\documentclass[11pt,a4paper]{article}

\usepackage{amssymb}
\usepackage[dvips]{graphicx}
\usepackage{bm}
\newcommand{\p}[1]{(\ref{#1})}
\def\theequation{\arabic{section}.\arabic{equation}}
\begin{document}

\title{Supergraph analysis of the one-loop divergences in $6D$, ${\cal N} = (1,0)$ and ${\cal N} = (1,1)$ gauge theories}

\author{I.L. Buchbinder\footnote{joseph@tspu.edu.ru}\\
{\small{\em Department of Theoretical Physics, Tomsk State Pedagogical
University,}}\\
{\small{\em 634061, Tomsk,  Russia}} \\
{\small{\em National Research Tomsk State University, 634050, Tomsk, Russia}},\\
\\
E.A. Ivanov\footnote{eivanov@theor.jinr.ru}\\
{\small{\em Bogoliubov Laboratory of Theoretical Physics, JINR, 141980 Dubna, Moscow region,
Russia}},\\
\\
B.S. Merzlikin\footnote{merzlikin@tspu.edu.ru}\\
{\small{\em Department of Theoretical Physics, Tomsk State Pedagogical
University}},\\
{\small{\em 634061, Tomsk,  Russia}}, \\
{\small{\em Department of Higher Mathematics and Mathematical Physics}},\\
{\small{\em \it Tomsk Polytechnic University, 634050, Tomsk, Russia}},\\
\\
K.V. Stepanyantz\footnote{stepan@m9com.ru}\\ {\small{\em Moscow State University}},\\
{\small{\em Faculty of Physics, Department of Theoretical Physics}},\\
{\small{\em 119991, Moscow, Russia}} }

\date{}

\maketitle

\begin{abstract}
We study the one-loop effective action for $6D,$ ${\cal
N}=(1,0)$ supersymmetric Yang--Mills (SYM) theory with hypermultiplets and $6D,$ ${\cal N}=(1,1)$ SYM theory as a subclass of the former,
using the off-shell formulation of
these theories in $6D,$ ${\cal N}=(1,0)$ harmonic superspace. We
develop the corresponding supergraph technique and apply it to compute
the one-loop divergences in the background field method ensuring the manifest gauge invariance. We calculate the two-point
Green functions of the gauge superfield and the hypermultiplet, as well as
the three-point gauge-hypermultipet Green function. Using these
Green functions and exploiting gauge invariance of the theory, we find the full set
of the off-shell one-loop divergent contributions, including the logarithmic and
power ones. Our results precisely match with those obtained earlier in \cite{Buchbinder:2016gmc,Buchbinder:2016url}
within the proper time superfield method.
\end{abstract}

\unitlength=1cm

\section{Introduction}
\hspace*{\parindent} Investigation of quantum corrections in
higher-dimensional gauge theories is an exciting problem with a long
history (see, e.g.,
\cite{Howe:1983jm,Howe:2002ui,Bossard:2009sy,Bossard:2009mn,Fradkin:1982kf,MarSag,Smilga:2016dpe,Bork:2015zaa}
and references therein). On the one hand, because of dimensionful
coupling constant, these theories are not renormalizable by formal
power-counting. On the other hand, extended supersymmetry is capable
to improve the ultraviolet behavior of a theory. And indeed, it was
shown in the above papers that, e.g., in six-dimensional
supersymmetric Yang-Mills theories the one- and two-loop amplitudes
are finite. It is extremely
interesting to analyze the impact of extended supersymmetry on a
general structure of ultraviolet divergences in higher-dimensional
gauge theories and to learn whether the supersymmetry  is  powerful
enough for construction of the renormalizable and, perhaps,  finite
higher dimensional quantum field-theoretical models.

To accomplish this program, it is natural to start with such a
formulation of the theory which makes manifest and off-shell as much
underlying symmetries as possible. In our case these are
supersymmetry and gauge invariance. In $4D$ theories, ${\cal N}=1$
supersymmetry becomes manifest in the ${\cal N}=1$ superfield
formalism (see e.g. \cite{Gates:1983nr,Buchbinder:1998qv}). Both
supersymmetries of $4D,$ ${\cal N}=2$ theories can also be made
manifest by making use of the ${\cal N}=2$ harmonic superspace
approach
\cite{Galperin:1984av,Galperin:1985bj,Galperin:1985va,Galperin:2001uw}.
The gauge invariance is manifest in the framework of background
field method, which can also be formulated in superspace.

In this paper we consider $6D,$ ${\cal N}=(1,0)$ and ${\cal
N}=(1,1)$ supersymmetric Yang--Mills (SYM) theories, which, to
certain extent, are similar to $4D,$ ${\cal N}=2$ and ${\cal N}=4$
SYM theories, respectively. {}From the ${\cal N}=(1,0)$
supersymmetry standpoint, such theories describe the interacting
gauge multiplet and hypermuptiplets. Both these theories can be
formulated in $6D,$ ${\cal N}=(1,0)$ harmonic superspace
\cite{Howe:1983fr,Howe:1985ar,Zupnik:1986da,Ivanov:2005qf,Ivanov:2005kz,Buchbinder:2014sna},
so that ${\cal N}=(1,0)$ supersymmetry remains a manifest off-shell
symmetry at all steps of quantum calculations. Moreover, the gauge
symmetry can be made manifest by using the background field method
which has been formulated in harmonic superspace in
\cite{Buchbinder:1997ya,Buchbinder:2001wy,Buchbinder:2016wng}. Thus,
the harmonic superspace approach augmented with the background field
method allows one to better figure out the restrictions imposed by
gauge symmetry and extended supersymmetry on the structure of the
ultraviolet divergences. However, it should be noted that, in
general, ${\cal N}=(1,0)$ theories are plagued by anomalies
\cite{Townsend:1983ana,Smilga:2006ax,Kuzenko:2015xiz} and it seems
impossible to construct a regularization which would simultaneously
preserve both supersymmetry and gauge symmetry. This is an essential
difference from the $4D$ case, where an invariant regularization for
${\cal N}=2$ supersymmetric gauge theories  can be constructed
\cite{Buchbinder:2014wra,Buchbinder:2015eva} as a proper
generalization of the higher-derivative regularization worked out in
\cite{Slavnov:1971aw,Slavnov:1972sq}.

Our basic aim in this paper is to study in detail an off-shell structure of the one-loop divergences of $6D,$
${\cal N}=(1,0)$ and ${\cal N}=(1,1)$ SYM theories, in both the gauge
multiplet and the hypermultiplet sectors.

Earlier in Refs. \cite{Buchbinder:2016gmc,Buchbinder:2016url} we
have studied the one-loop divergences using the operator proper --
time method in ${\cal N}=(1,0)$ harmonic superspace (for the case of
non-supersymmetric theories this method was initiated in
\cite{Schwinger:1951nm,DeWitt:1965jb}). It has been demonstrated
that the general ${\cal N}=(1,0)$ theory with hypermultiplets in an
arbitrary representation $R$ of the gauge group $G$ is divergent in
the one-loop approximation. However, in the special case of ${\cal
N}=(1,1)$ SYM theory, which corresponds to the hypermultiplet in the
adjoint representation the divergences cancel each other and the
theory proves to be one-loop finite off shell. It gave us a ground
to expect a better ultraviolet  behavior of this theory in higher
loops as well. It is worth pointing out that the $4D$ analog of
${\cal N}=(1,1)$ theory is ${\cal N}=4$ SYM theory, which is
finite to all loops
\cite{Grisaru:1982zh,Mandelstam:1982cb,Brink:1982pd,Howe:1983sr}.

In this paper we develop in detail the harmonic supergraph approach
to the study of the one-loop divergences in $6D,$ ${\cal N}=(1,0)$ and
${\cal N}=(1,1)$ SYM theories. Such an approach for calculating the
structure of divergences is more familiar, as compared to the operator proper-time method, and it provides an appropriate basis for studying the higher-loop
divergences. Besides, we will clarify and justify some subtle aspects of
the calculations which have been performed in our previous papers
\cite{Buchbinder:2016gmc,Buchbinder:2016url}.
%in too short form.

The proper-time technique is very efficient for one-loop
calculations. However, for calculating the higher-loop contributions
to effective action this technique turns out not so convenient. Usually, for calculation of
the divergent diagrams, a simpler technique is used. Just such a technique is developed in this
paper, with a possibility of its further applications for higher-loop
calculations. Note that in the ${\cal N}=(1,0)$ harmonic superspace approach the number of divergent
one-loop supergraphs is infinite, because the gauge superfield is
dimensionless. Surely, it is difficult to calculate exactly a sum
of infinite number of divergent supergraphs. However, it is
possible to calculate divergent diagrams with small numbers of
external gauge lines and then to restore the exact result by gauge symmetry
arguments. In this paper we demonstrate how this method can be
applied for calculating the divergent part of the one-loop effective
action of ${\cal N}=(1,0)$ SYM theory with the hypermultiplet in an
arbitrary representation of the gauge group.

The paper is organized as follows. In Sect. \ref{Section_Theory} we
formulate ${\cal N}=(1,0)$ SYM theory with hypermultiplets in harmonic superspace and
introduce the notation. Sect. \ref{Section_Quantization} is
devoted to the quantization of the theory. In particular, we
construct the background field method and describe the gauge
fixing procedure. Feynman rules for the theory under
consideration are presented in Sect. \ref{Section_Feynman_Rules}.
Using these rules, in Sect. \ref{Section_Divergences} we
calculate the divergent supergraphs with the minimal numbers of external
gauge legs and then restore the full result for the divergent part of the
one-loop effective action by the gauge symmetry reasonings.
The results obtained are listed and discussed in Sect. \ref{Section_Summary}. Technical details of
the harmonic supergraph calculations are collected in Appendices A and B.

\section{${\cal N}=(1,0)$ supersymmetric gauge theories in $6D$ harmonic \break superspace}
\hspace*{\parindent}\label{Section_Theory}

The harmonic superspace approach \cite{Galperin:2001uw} is convenient for describing extended supersymmetric theories, mainly because all symmetries of the theory
in this approach  are manifest. In our notation the harmonic variables are denoted by $u^{\pm i}$, where $u_i^- = (u^{+i})^*$. These variables are constrained by
the condition $u^{+i} u_i^- = 1$. The anticommuting left-handed spinor coordinates are denoted by $\theta^{a}_i$ and the usual coordinates are denoted by $x^M$,
where $M=0,\ldots 5$.
The coordinates of the ordinary ${\cal N}=2$ superspace are $z\equiv (x^M,\theta^a_i)$, and $\zeta\equiv (x^M_A,\theta^{+a})$ are analytic coordinates
defined as

\begin{equation}
x^M_A \equiv x^M + \frac{i}{2}\theta^{-}\gamma^M \theta^+;\qquad \theta^{\pm a} \equiv u^\pm_i \theta^{ai},
\end{equation}

\noindent
where $\gamma^M$ are six-dimensional $\gamma$-matrices. This implies that the corresponding integration measures can be written as

\begin{equation}\label{Integrations}
\int d^{14}z = \int d^6x\,d^8\theta;\qquad \int d\zeta^{(-4)} \equiv \int d^6x\, d^4\theta^+.
\end{equation}

\noindent
Note that

\begin{equation}\label{Measure}
\int d^8\theta = \int d^4\theta^{+} (D^+)^4,
\end{equation}

\noindent
where we have introduced the notation

\begin{equation}
(D^+)^4 = -\frac{1}{24}\varepsilon^{abcd} D_a^+ D_b^+ D_c^+ D_d^+
\end{equation}

\noindent
with $D^+_a = u^{+}_i D_{a}^i$ (similarly, $D^-_a \equiv u^{-}_i D_{a}^i$).

In harmonic superspace the action of the $6D, {\cal N}=(1,0)$  SYM theory has the form \cite{Zupnik:1986da}

\begin{equation}\label{Action_N2SYM}
S_{\mbox{\scriptsize SYM}} = \frac{1}{f_0^2} \sum\limits_{n=2}^\infty \frac{(-i)^{n}}{n} \mbox{tr} \int d^{14}z\, du_1 \ldots du_n\,
\frac{V^{++}(z,u_1)\ldots V^{++}(z,u_n)}{(u_1^+ u_2^+) \ldots (u_n^+ u_1^+)}\,,
\end{equation}

\noindent
where $f_0$ is the bare coupling constant, which in $6D$ has the dimension $m^{-1}$. The gauge superfield $V^{++}(z,u)$ satisfies the analyticity condition

\begin{equation}
D^+_a V^{++} = 0
\end{equation}

\noindent
and is real with respect to the special ``tilde'' conjugation, $\,\widetilde{V^{++}} = V^{++}\,$\,. It can be presented as
$V^{++}(z,u) = V^{++A} t^A$, where $t^A$ are the generators of the fundamental representation of the gauge group $G$.
In our notation they satisfy the relations

\begin{equation}
\mbox{tr}(t^A t^B) = \frac{1}{2}\delta^{AB};\qquad [t^A,t^B] = if^{ABC} t^C,
\end{equation}

\noindent
where $f^{ABC}$ are the gauge group structure constants.

The expression for the SYM action is essentially simplified in the abelian case. Namely, only terms quadratic in the gauge superfield $V^{++}$
survive:

\begin{equation}\label{Action_Gauge}
S_{U(1)} = \frac{1}{4f_0^2} \int d^{14}z\,\frac{du_1 du_2}{(u_1^+ u_2^+)^2} V^{++}(z,u_1) V^{++}(z,u_2).
\end{equation}

\noindent

In this paper we will consider $6D,$ ${\cal N}=(1,0)$ SYM theory with massless hypermultiplets residing in a certain representation $R$ of gauge group.
In the harmonic superspace formalism, the total action of such a system reads

\begin{eqnarray}\label{Action}
S = \frac{1}{f_0^2} \sum\limits_{n=2}^\infty \frac{(-i)^{n}}{n} \mbox{tr} \int d^{14}z\, du_1 \ldots du_n\,
\frac{V^{++}(z,u_1)\ldots V^{++}(z,u_n)}{(u_1^+ u_2^+) \ldots (u_n^+ u_1^+)} - \int d\zeta^{(-4)} du\,\widetilde q^+ \nabla^{++} q^+\,,
\end{eqnarray}

\noindent
where the analytic superfield $q^+$ describes the hypermultiplet. The covariant harmonic derivative in Eq. (\ref{Action}) is defined as

\begin{equation}
\nabla^{++} = D^{++} + i V^{++} = D^{++} + i V^{++A} T^A\,. \label{Cov++}
\end{equation}

\noindent
The ``flat'' harmonic derivatives $D^{\pm\pm}, D^0$ are defined by\footnote{One can easily see that they form an $SU(2)$ algebra.}

\begin{equation}
D^{++} = u^{+i} \frac{\partial}{\partial u^{-i}};\qquad D^{--} = u^{-i} \frac{\partial}{\partial u^{+i}};\qquad
D^0 = u^{+i} \frac{\partial}{\partial u^{+i}} - u^{-i} \frac{\partial}{\partial u^{-i}}
\end{equation}

\noindent
and $T^A$ in (\ref{Cov++}) are the generators of the gauge group in the representation $R$, such that $[T^A,T^B]=if^{ABC} T^C$. We will consider only simple gauge groups,
so that

\begin{equation}
\mbox{tr}(T^A T^B) = T(R)\delta^{AB};\quad \mbox{tr}(T_{\mbox{\scriptsize Adj}}^A T_{\mbox{\scriptsize Adj}}^B) = f^{ACD} f^{BCD} = C_2\delta^{AB};
\quad (T^A T^A)_i{}^j = C(R)_i{}^j.
\end{equation}

\noindent
The representation $R$ can in general be  reducible. For an irreducible representation $R$, $C(R)_i{}^j$ is proportional to $\delta_i^j$. If $R$
is the adjoint representation, the action (\ref{Action}) describes the ${\cal N}=(1,1)$ SYM theory. In this case, the action (\ref{Action}) is invariant under
an extra hidden ${\cal N}=(0,1)$ supersymmetry which mixes the gauge superfield and the hypermultiplet.

The ${\cal N}=(1,1)$ SYM action  (\ref{Action}) is invariant under the gauge transformations

\begin{equation}\label{Gauge_Transformations}
V^{++} \to  e^{i\lambda} V^{++} e^{-i\lambda}  - i e^{i\lambda} D^{++}e^{-i\lambda}; \qquad  q^+ \to  e^{i\lambda} q^+,
\end{equation}

\noindent where $\lambda = \lambda^A t^A\,$, when checking the invariance of the pure gauge-field part
of the action, and $\lambda = \lambda^A T^A$ while dealing with the hypermultiplet
part. The parameters $\lambda^A$ are the analytic superfields which are real with
respect to the tilde-conjugation, $\widetilde{\lambda^A} =
\lambda^A$\,.

One more necessary ingredient of the superfield formalism is a non-analytic superfield $V^{--}$ introduced as a solution of the ``harmonic flatness condition''

\begin{equation}\label{V--Equation}
D^{++} V^{--} - D^{--} V^{++} +i[V^{++}, V^{--}]=0\,.
\end{equation}

\noindent
This superfield  can be solved for from (\ref{V--Equation}) in terms of $V^{++}$   as

\begin{equation}\label{V--_Definition}
V^{--}(z,u) \equiv \sum\limits_{n=1}^\infty (-i)^{n+1}\int du_1\,\ldots\,du_n\,\frac{V^{++}(z,u_1) \ldots V^{++}(z,u_n)}{(u^+ u_1^+)(u_1^+u_2^+)\ldots (u_n^+ u^+)}.
\end{equation}

\noindent
Under gauge transformations (\ref{Gauge_Transformations}) $V^{--}$ is transformed as

\begin{equation}
V^{--} \to  e^{i\lambda} V^{--} e^{-i\lambda}  - i e^{i\lambda} D^{--}e^{-i\lambda}.
\end{equation}

\noindent
{}From the geometric point of view, this superfield is the connection covariantizing the harmonic derivative $D^{--}$:

\begin{equation}
D^{--} \;\Rightarrow \;\nabla^{--} \equiv D^{--} + i V^{--}.
\end{equation}

\noindent
It can be used to construct  the important analytic superfield strength

\begin{equation}
F^{++} \equiv (D^+)^4 V^{--},
\end{equation}

\noindent
which transforms homogeneously, as $F^{++} \to e^{i\lambda} F^{++} e^{-i\lambda}$.

In the abelian case, the action of the gauge theory-hypermultiplet system can be written as

\begin{eqnarray}\label{Action_Abelian}
S = \frac{1}{4f_0^2} \int d^{14}z\,\frac{du_1 du_2}{(u_1^+ u_2^+)^2} V^{++}(z,u_1) V^{++}(z,u_2) - \int d\zeta^{(-4)} du\,\widetilde q^+ \nabla^{++} q^+,
\end{eqnarray}

\noindent
where $\nabla^{++} = D^{++} + i V^{++}$, and it is invariant under the gauge transformations

\begin{equation}\label{Gauge_Transformations_Abelian}
V^{++} \to  V^{++}  - D^{++}\lambda; \qquad V^{--} \to  V^{--}  - D^{--}\lambda; \qquad  q^+ \to  e^{i\lambda} q^+.
\end{equation}

\noindent
In this case Eq. (\ref{V--Equation}) becomes linear,

\begin{equation}
D^{++} V^{--} = D^{--} V^{++},
\end{equation}

\noindent
and so provides  the linear solution for $V^{--}\,$,

\begin{equation}\label{V--_Abelian}
V^{--}(z,u) = \int du_1\,\frac{V^{++}(z,u_1)}{(u^+ u_1^+)^2}.
\end{equation}

\noindent
The analytic superfield $F^{++}$ in the abelian case is gauge invariant.
\setcounter{equation}{0}

\section{Quantization and background field method in ${\cal N}=(1,0)$ \break harmonic superspace}
\hspace*{\parindent}\label{Section_Quantization}

It is convenient to quantize $6D$, ${\cal N}=(1,0)$ theories directly in ${\cal N}=(1,0)$ harmonic superspace,
thus ensuring the manifestly supersymmetric form of the quantum corrections. It is also convenient to make use of
the background field method, which gives the manifestly gauge invariant effective action.

The background-quantum splitting is introduced by the substitution

\begin{equation}
V^{++} = \bm{V}^{++} + v^{++},
\end{equation}

\noindent
where $\bm{V}^{++}$ denotes the background gauge superfield, while $v^{++}$ is the quantum gauge superfield. In supergraphs,
external lines corresponding to the background gauge superfield will be denoted by the bold wavy lines, while the external lines
corresponding to the quantum gauge superfield - by the usual wavy lines.
Note that we do not make the background-quantum splitting for the hypermultiplet. This is admissible because such a splitting is linear
and we will choose the gauge-fixing term to be independent of the hypermultipet. This implies that the effective action will depend
only on the sum of the quantum and background hypermultiplet superfields, so that there is no actual need to separately introduce
the background hypermultiplet superfield.

After the background-quantum splitting, gauge invariance (\ref{Gauge_Transformations}) amounts to the background gauge invariance

\begin{equation}\label{Background_Gauge_Transformations}
\bm{V}^{++} \to  e^{i\lambda} \bm{V}^{++} e^{-i\lambda}  - i e^{i\lambda} D^{++}e^{-i\lambda}; \qquad v^{++} \to e^{i\lambda} v^{++} e^{-i\lambda};\qquad
q^+ \to  e^{i\lambda} q^+.
\end{equation}

\noindent
and the quantum gauge invariance

\begin{equation}\label{Quantum_Gauge_Transformations}
\bm{V}^{++} \to  e^{i\lambda}\bm{V}^{++} e^{-i\lambda}; \qquad v^{++} \to e^{i\lambda} v^{++} e^{-i\lambda}
- i e^{i\lambda} D^{++}e^{-i\lambda};\qquad  q^+ \to  e^{i\lambda} q^+.
\end{equation}

\noindent
To obtain the gauge invariant effective action, one should fix a gauge only with respect to the quantum superfields,
without breaking the background gauge invariance (\ref{Background_Gauge_Transformations}). For example, it is possible to add the following gauge-fixing term:

\begin{eqnarray}\label{GF_Term}
&& S_{\mbox{\scriptsize gf}} = - \frac{1}{2f_0^2\xi_0}\mbox{tr}\int d^{14}z\, du_1 du_2 \frac{(u_1^- u_2^-)}{(u_1^+ u_2^+)^3} D_1^{++} \Big[e^{-i\bm{b}(z,u_1)}
v^{++}(z,u_1) e^{i\bm{b}(z,u_1)}\Big] \nonumber\\
&& \times D_2^{++} \Big[e^{-i\bm{b}(z,u_2)}v^{++}(z,u_2)e^{i\bm{b}(z,u_2)}\Big],
\end{eqnarray}

\noindent
where $\bm{b}(z,u)$ is the background bridge superfield. The bridge is related to the background superfields $\bm{V}^{++}$ and $\bm{V}^{--}$ by the relations

\begin{equation}
\bm{V}^{++} = - i e^{i\bm{b}} D^{++} e^{-i\bm{b}};\qquad \bm{V}^{--} = - i e^{i\bm{b}} D^{--} e^{-i\bm{b}}.
\end{equation}

\noindent
Note that the hypermultiplet does not enter Eq. (\ref{GF_Term}) and the theory is invariant under the background gauge transformations
even without the background-quantum splitting for the $q^+$ superfields.

The expression (\ref{GF_Term}) is an analog of the usual $\xi$-gauge. The terms quadratic in the quantum gauge superfield in the total action become

\begin{eqnarray}
&& S^{(2)} + S_{\mbox{\scriptsize gf}}^{(2)} = \frac{1}{2f_0^2}\Big(1-\frac{1}{\xi_0}\Big)\mbox{tr}\int d^{14}z\, du_1 du_2 \frac{1}{(u_1^+ u_2^+)^2} v^{++}(z,u_1) v^{++}(z,u_2) \nonumber\\
&& +\, \frac{1}{2f_0^2\xi_0}\mbox{tr}\int d\zeta^{(-4)}\, du\, v^{++}(z,u) \Box v^{++}(z,u), \qquad \label{Quadr}
\end{eqnarray}

\noindent
where $\Box\equiv \partial^2$. We observe that in the Feynman gauge $\xi_0=1$ this expression is essentially simplified, as in the case of the usual Yang--Mills theory.
Note that, when deriving (\ref{Quadr}), we integrated by parts and used the relation

\begin{equation}\label{U_Identity}
D_1^{++} \frac{1}{(u_1^+ u_2^+)^3} = \frac{1}{2} (D_1^{--})^2 \delta^{(3,-3)}(u_1,u_2).
\end{equation}

\noindent
Then we again integrated by parts with respect to $(D_1^{--})^2$ , taking into account Eq. (\ref{Measure}) and the identity

\begin{equation}
\frac{1}{2} (D^+)^4 (D^{--})^2 v^{++} = \Box v^{++},
\end{equation}

\noindent
which follows from the analyticity of the quantum gauge superfield $v^{++}$.

The gauge-fixing term (\ref{GF_Term}) is invariant under the transformations (\ref{Background_Gauge_Transformations})
which act on the bridge as

\begin{equation}
e^{i\bm{b}} \to e^{i\lambda} e^{i\bm{b}} e^{i\tau},
\end{equation}

\noindent
where $\tau=\tau(x,\theta)$ is a gauge parameter independent of the harmonic variables.

Note that in the abelian case the bridge superfield is not present in the gauge-fixing action. Also, in the abelian case
it is not necessary to introduce ghosts. The latter are required by the quantization procedure only for non-abelian gauge theories.
In the considered theory one is led to insert, into the generating functional, some determinants corresponding to the Faddeev--Popov and Nielsen--Kallosh ghosts.

The Faddeev--Popov ghosts $b$ and $c$ are anticommuting analytic superfields in the adjoint representation of the gauge group. The action for them has the form

\begin{equation}\label{Action_Faddeev_Popov}
S_{\mbox{\scriptsize FP}} = \mbox{tr} \int d\zeta^{(-4)}\, du\, b \bm{\nabla}^{++}\Big( \bm{\nabla}^{++} c + i[v^{++},c]\Big),
\end{equation}

\noindent
where $\bm{\nabla}^{++} c = D^{++} c + i [\bm{V}^{++}, c]$ is the background covariant derivative.

Also it is necessary to insert into the generating functional the determinants corresponding to the Nielsen--Kallosh ghosts,

\begin{equation}\label{Action_Nielsev_Kallosh}
\Delta_{NK} \equiv \mbox{Det}^{1/2} \stackrel{\bm{\frown}}{\bm{\Box}} \int D\varphi \exp\big(i S_{\mbox{\scriptsize NK}}\big),
\end{equation}

\noindent
where $\stackrel{\bm{\frown}}{\bm{\Box}}\equiv \frac{1}{2} (D^+)^4 (\bm{\nabla}^{--})^2$ and $\varphi$
is a commuting analytic Nielsen--Kallosh ghost superfield in the adjoint representation of the gauge group with

\begin{equation}
S_{\mbox{\scriptsize NK}} = -\frac{1}{2} \mbox{tr} \int d\zeta^{(-4)}\, du\, (\bm{\nabla}^{++}\varphi)^2.
\end{equation}

\noindent
Introducing anticommuting analytic superfields $\xi^{(+4)}$ and $\sigma$ in the adjoint representation of the gauge group,
one can write  $\mbox{Det} \stackrel{\bm{\frown}}{\bm{\Box}}$ in the form of the functional integral,

\begin{equation}\label{Determinant}
\mbox{Det} \stackrel{\bm{\frown}}{\bm{\Box}} = \int D\xi^{(+4)} D\sigma \exp\Big(i\,\mbox{tr}\int d\zeta^{(-4)}\,du\,\xi^{(+4)}
\stackrel{\bm{\frown}}{\bm{\Box}} \sigma \Big).
\end{equation}

Thus, the generating functional of the considered theory can be written as

\begin{equation}
Z = \int Dv^{++}\,D\widetilde q^+\, Dq^+\,Db\,Dc\,D\varphi\,\mbox{Det}^{1/2} \stackrel{\bm{\frown}}{\bm{\Box}} \exp\Big[i(S+S_{\mbox{\scriptsize gf}}
+S_{\mbox{\scriptsize FP}} + S_{\mbox{\scriptsize NK}}+ S_{\mbox{\scriptsize sources}})\Big],
\end{equation}

\noindent
where $S_{\mbox{\scriptsize sources}}$ denotes the relevant source terms. For example, the sources for the quantum gauge and hypermultiplet superfields
can be introduced as

\begin{equation}\label{Sources}
\int d\zeta^{(-4)}\,du\, \Big[v^{++ A} J^{(+2)A} + j^{(+3)i} (q^+)_i + \widetilde j^{(+3)}_i (\widetilde q^+)^i\Big].
\end{equation}

\noindent
Likewise, it is possible to add sources for other  superfields involved.
\setcounter{equation}{0}

\section{Propagators, vertices and supergraphs}
\hspace*{\parindent}\label{Section_Feynman_Rules}

The Feynman rules for $6D,$ ${\cal N}=(1,0)$ supersymmetric gauge theories in harmonic superspace are very similar to those in the case of $4D,$ ${\cal N}=2$ supersymmetric
theories which have been considered in detail in \cite{Galperin:1985bj,Galperin:1985va}.

In order to find the propagator of the quantum gauge superfield, we consider the linearized equation of motion for this superfield (setting the background gauge field
equal to zero) in the presence of the source term (\ref{Sources}):

\begin{equation}
\frac{1}{2\xi_0 f_0^2} \Box v^{++A}(z,u_1) + \frac{1}{2f_0^2}\Big(1-\frac{1}{\xi_0}\Big) \int du_2 \frac{1}{(u_1^+ u_2^+)^2} (D_1^+)^4 v^{++A}(z,u_2) + J^{(+2)A}(z,u_1) = 0.
\end{equation}

\noindent
Its solution can be written as

\begin{equation}
v^{++A}(z,u_1) = -\frac{2\xi_0 f_0^2}{\Box} J^{(+2)A}(z,u_1) + \frac{2f_0^2(\xi_0-1)}{\Box^2} \int du_2 \frac{1}{(u_1^+ u_2^+)^2} (D_1^+)^4 J^{(+2)A}(z,u_2).
\end{equation}

\noindent
This implies that the propagator of the gauge superfield in the $\xi$-gauge reads

\begin{eqnarray}\label{Gauge_Propagator}
&& (G_V^{(2,2)})^{AB}(z_1,u_1;z_2,u_2) = - 2 f_0^2 \Big(\frac{\xi_0}{\Box} (D_1^+)^4 \delta^{(2,-2)}(u_2,u_1)\nonumber\\
&&\qquad\qquad\qquad\qquad\qquad - \frac{\xi_0-1}{\Box^2} (D_1^+)^4 (D_2^+)^4 \frac{1}{(u_1^+ u_2^+)^2}\Big) \delta^6(x_1-x_2) \delta^8(\theta_1-\theta_2)\delta^{AB}.\qquad
\end{eqnarray}

\begin{figure}[h]
\begin{picture}(0,2)
\put(1,0.2){\includegraphics[scale=0.2]{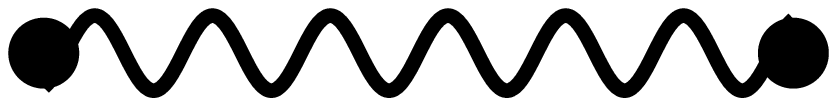}}
\put(0.5,1){(1)}
\put(5,0.2){\includegraphics[scale=0.2]{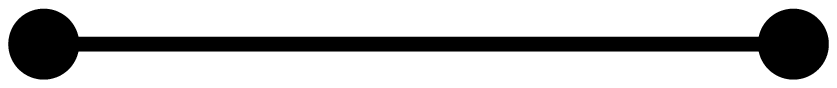}}
\put(4.5,1){(2)}
\put(9,0.2){\includegraphics[scale=0.2]{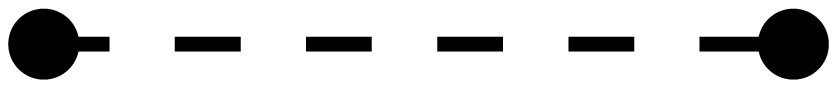}}
\put(8.5,1){(3)}
\put(13,0.2){\includegraphics[scale=0.2]{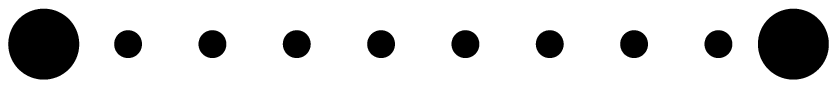}}
\put(12.5,1){(4)}
\end{picture}
\caption{Propagators of various superfields: (1), (2), (3), and (4) stand for the gauge, hypermultiplet, Faddeev--Popov and Nielsen--Kallosh ghost propagators,
respectively.}\label{Figure_Propagators}
\end{figure}

\noindent
Below we will use the gauge $\xi_0=1$, in which the propagator has the simplest form, with the second term vanishing. The propagator of the gauge superfield
will be graphically represented by the wavy line ending at the points 1 and 2, see Fig. \ref{Figure_Propagators}, where it is denoted by (1).

The propagator of the hypermultiplet superfields can be defined similarly, and it is given by the expression

\begin{equation}\label{Hypermultiplet_Propagator}
(G_q^{(1,1)})_i{}^j(z_1,u_1;z_2,u_2) = (D_1^+)^4 (D_2^+)^4 \frac{1}{\Box} \delta^{14}(z_1-z_2) \frac{1}{(u_1^+ u_2^+)^3}\delta_i{}^j\,,
\end{equation}

\noindent
where

\begin{equation}
\delta^{14}(z_1-z_2) \equiv \delta^6(x_1-x_2) \delta^8(\theta_1-\theta_2).
\end{equation}

\noindent
This propagator will be represented by the line ending at the points 1 and 2. It is denoted by the symbol (2) in Fig. \ref{Figure_Propagators}.
The external hypermultiplets will be also denoted by such lines.

The propagators of the Faddeev--Popov and Nielsen--Kallosh ghosts have the form

\begin{equation}
\frac{(D_1^+)^4 (D_2^+)^4}{2\Box}
\delta^{14}(z_1-z_2) \frac{(u_1^- u_2^-)}{(u_1^+ u_2^+)^3}\delta^{AB}.
\end{equation}

\noindent
They are given, respectively, by the dashed and dotted lines connecting the points 1 and 2 and denoted in Fig. \ref{Figure_Propagators} by the symbols (3) and (4).

Also, we will need the propagator of the superfields $\xi^{(+4)}$ and $\sigma$ introduced in Eq. (\ref{Determinant}). It is easy to see that it is given by the expression

\begin{equation}
-\frac{(D_1^+)^4}{2\Box} \delta^{14}(z_1-z_2) \delta^{(0,0)}(u_1,u_2)\delta^{AB}.
\end{equation}

\begin{figure}[h]
\begin{picture}(0,2)
\put(4,0){\includegraphics[scale=0.15]{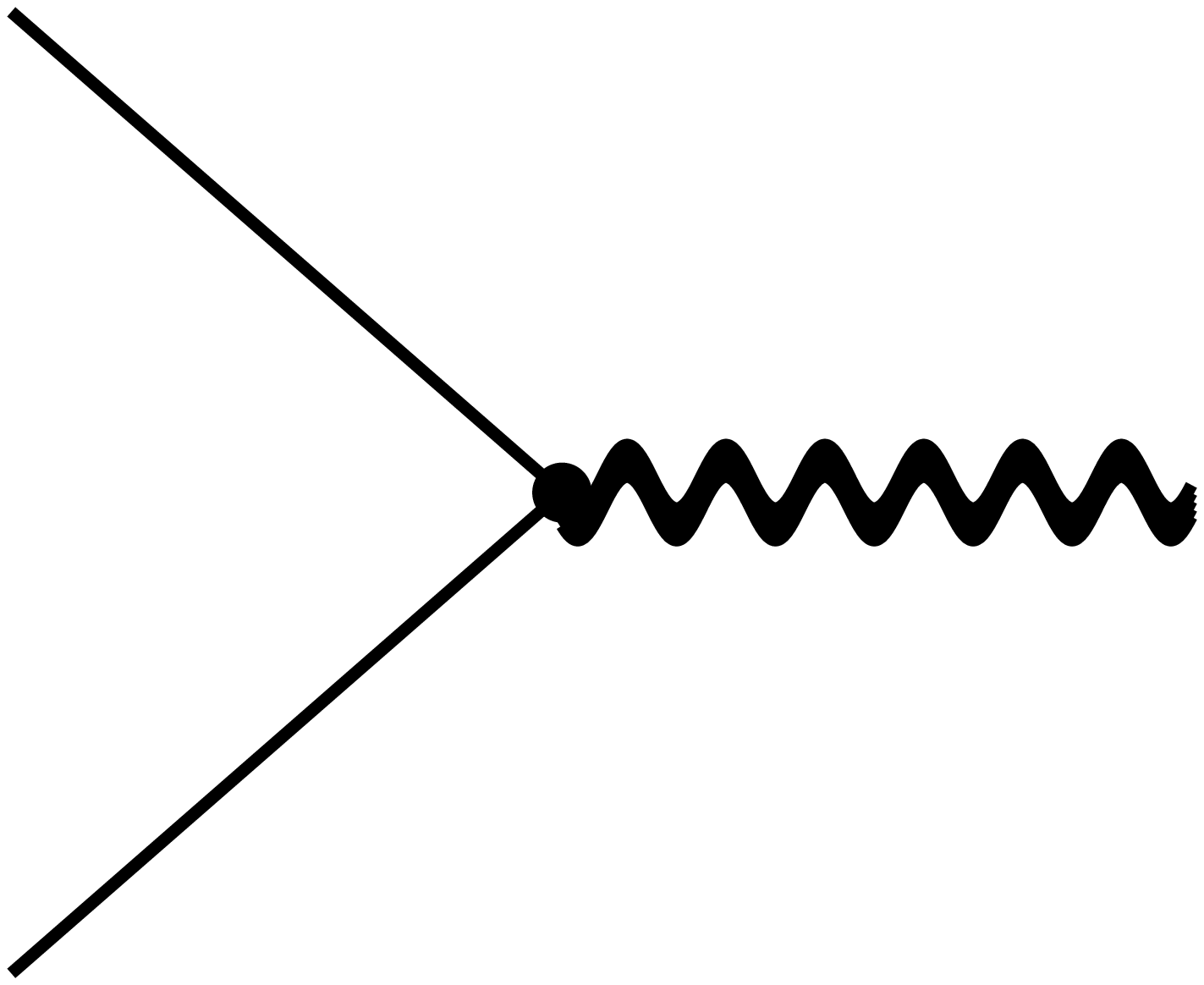}}
\put(10,0){\includegraphics[scale=0.15]{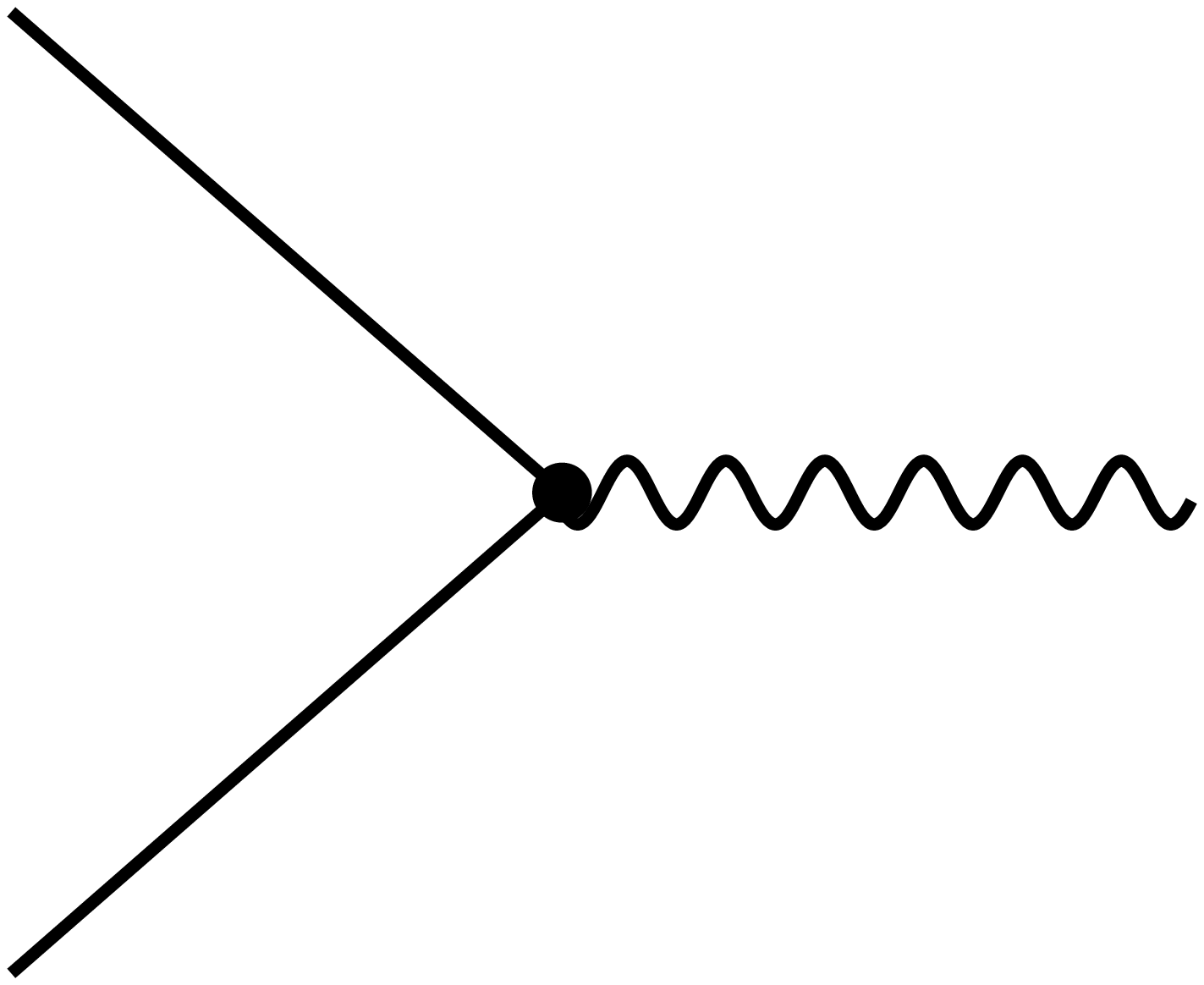}}
\end{picture}
\caption{Vertices coming from the hypermultiplet part of the action.}
\label{Figure_Vertex}
\end{figure}

The only coupling of the gauge superfield with the hypermultiplet ones comes from the hypermultiplet action:

\begin{equation}
S_I = - i \int d\zeta^{(-4)}\, du\, (\widetilde q^{+})^i (V^{++})_i{}^j (q^+)_j = - i \int d\zeta^{(-4)}\, du\, (\widetilde q^{+})^i (\bm{V}^{++} + v^{++})_i{}^j (q^+)_j,
\end{equation}

\noindent
where $(V^{++})_i{}^j = f_0 V^{++A} (T^A)_i{}^j$. This implies that the supergraphs can contain vertices with two hypermultiplet legs and one leg
of the background or quantum gauge superfield. These vertices are presented in Fig. \ref{Figure_Vertex}. In the abelian case, these are the only interaction vertices at all.

In the non-abelian case the are infinitely many interaction vertices, because the action (\ref{Action_N2SYM}) contains terms involving all powers
($n\ge 2$) of the gauge superfield $V^{++}$. Clearly, each line in such vertices represents the quantum superfield $v^{++}$ or the background gauge superfield $\bm{V}^{++}$.
Expressions for the vertices with purely quantum gauge legs can be read off from Eq. (\ref{Action_N2SYM}). The vertices with legs of the background gauge superfield
can also come from the gauge fixing action (\ref{GF_Term}).  On the external background superfield legs in such vertices there always appears the bridge superfield.

As the action (\ref{Action_Faddeev_Popov}) contains two background supersymmetric derivatives $\bm{\nabla}^{++}$, there are vertices with two Faddeev--Popov ghost
legs and one or two legs of the gauge superfield. Note that the maximal number of the quantum gauge superfield legs equals one, because the Faddeev--Popov action
contains only first degree of $v^{++}\,$.

The Nielsen--Kallosh ghosts interact only with the background gauge superfield. Like in the case of Faddeev--Popov ghosts, only three- and four-point vertices are present.
The vertices containing legs of the superfields $\xi^{(+4)}$ and $\sigma$ can involve an arbitrary number of the external background superfield $\bm{V}^{++}$ legs,
because the operator $\stackrel{\bm{\frown}}{\bm{\Box}}$ contains the superfield $\bm{V}^{--}$ given by an infinite series  (\ref{V--_Definition}).
\setcounter{equation}{0}

\section{Structure of one-loop divergences}
\label{Section_Divergences}

\subsection{General analysis}
\hspace*{\parindent}

According to the general analysis performed in \cite{Bossard:2015dva} the
on-shell logarithmic divergences in the one-loop approximation can
be written as

\begin{equation}\label{Divergences}
\Gamma^{(1)}_{\infty,\ln} = \int d\zeta^{(-4)}\,du\,\Big[c_1 (F^{++ A})^2 +
i c_2  F^{++ A} (\widetilde q^+)^i (T^A)_i{}^j (q^+)_j + c_3 \Big((\widetilde q^{+})^i (q^+)_i\Big)^2\Big],
\end{equation}

\noindent where $c_1$, $c_2$, and $c_3$ are numerical real
coefficients. They have been found in \cite{Buchbinder:2016gmc}
by using the proper time method.

The degree of divergence in ${\cal N}=(1,0)$ gauge theory can be
deduced as follows. The effective action is dimensionless. On the other hand,
any contribution to this dimensionless effective action can be
presented as an integral over the total superspace. The harmonic
variables are dimensionless, while $[d^6x] = m^{-6}$ and
$[d^8\theta] = m^4$. The gauge superfields on the external legs are
dimensionless, $[V^{++}]=m^0$, while the external hypermultiplet legs
contribute $[q^+]= m^2$. In our notation, each gauge propagator gives
$f_0^2$, where $[f_0]=m^{-1}$, and each purely gauge vertex gives
$f_0^{-2}$. If $N_D$ spinor derivatives act to the external lines,
they also contribute $N_D/2$ to the total dimension. Therefore,
taking into account that the effective action is dimensionless, we
obtain that the dimension of the momentum integral $\omega$ in a
supergraph with $P_{V}$ gauge propagators, $V_V$ purely gauge
vertices, and $N_{q}$ external hypermultiplet legs should be equal
to

\begin{equation}
\omega = 6-4 + 2P_{V} - 2V_V - 2N_{q} - \frac{1}{2}N_D\,.
\end{equation}

\noindent The quantity $\omega$ is the superficial degree of
divergence. For hypermultiplets the number of external legs can be
written as $N_{q} = 2(-P_{q} + V_q)$, where $P_q$ and $V_q$ are
numbers of hypermultiplet propagators and the hypermultiplet-containing
vertices, respectively. For the closed loops of the Faddeev--Popov
ghosts the similar equality is $P_{\mbox{\scriptsize FP}} =
V_{\mbox{\scriptsize FP}}$, where $P_{\mbox{\scriptsize FP}}$ and
$V_{\mbox{\scriptsize FP}}$ are numbers of the Faddeev--Popov ghost
propagators and vertices, respectively. Using these relations  we
obtain

\begin{eqnarray}
\omega &=& 2 + 2(P_{V} + P_{q} + P_{\mbox{\scriptsize FP}}) - 2 (V_{q}+ V_V + V_{\mbox{\scriptsize FP}}) - N_{q} -\frac{1}{2}N_D\nonumber\\
&=& 2 - 2V + 2P - N_{q}-\frac{1}{2}N_D\,,
\end{eqnarray}

\noindent
where

\begin{equation}
P = P_{V} + P_{q} + P_{\mbox{\scriptsize FP}}\qquad \mbox{and}\qquad V = V_{V} + V_{q} + V_{\mbox{\scriptsize FP}}
\end{equation}

\noindent are total numbers of propagators and vertices in the
considered diagram, respectively. Since the number of loops is $L
= 1-V+P$, the result for the degree of divergence can be also rewritten as

\begin{equation}\label{Divergence_Degree}
\omega = 2L - N_q - \frac{1}{2} N_D.
\end{equation}

The supergraph is convergent if $\omega <0$, otherwise it is
divergent. The relation (\ref{Divergence_Degree}) allows one to list all
possible types of divergent supergraphs and compare the
corresponding counterterms with expression (\ref{Divergences}). In
the one-loop approximation the divergences correspond to $\omega$=2
and 0. Further analysis depends on the choice of regularization.

Let $N_q=0$, $N_D=0$, so that $\omega$=2, and use the dimensional-reduction type
of regularization. Then the only admissible counterterm in the gauge
multiplet sector is given by the first term in (\ref{Divergences}), with
dimensionless divergent coefficient $c_1$. Being dimensionless, this coefficient must be proportional to $1/\varepsilon\,,$
where ${\varepsilon}=d-6$ is a regularization parameter. Let still
$N_q=0$, $N_D=0$, $\omega$=2, but use now the cut-off
regularization. In this case we have the cut-off momentum
${\Lambda}$ and, hence, there are two admissible counterterms in
the gauge multiplet sector. Like in the previous case, one of them is given
by the first term in (\ref{Divergences}) with dimensionless divergent
coefficient $c_1$, which should now be proportional to
$\ln{\Lambda}$. The second one is proportional to ${\Lambda}^2$
multiplied by the classical action of the pure $6D$, ${\cal N}=(1,0)$ SYM
theory.

Let $N_q=2$, $N_D=0$ (so that $\omega=0$) and use the dimensional-reduction
regularization. The admissible counterterm is given by the second term
in (\ref{Divergences}) with the dimensionless divergent coefficient
$c_2$, which must be proportional to $1/\varepsilon$. In the case of the cut-off regularization we again obtain the counterterm
corresponding to the second term in (\ref{Divergences}) with
dimensionless divergent coefficient $c_2$ proportional to $\ln{\Lambda}$. Also, from Eq.
(\ref{Divergence_Degree}) we derive that $c_3=0$, because the
relevant structure corresponds to the convergent graphs with
$N_q=4$, $N_D=0$ and, hence, $\omega = -2\,$.

In this paper we carry out the one-loop calculations both in the
dimensional reduction scheme and in the cut-off regularization. We
confirm the results of \cite{Buchbinder:2016gmc} by an independent
calculation of superdiagrams in the dimensional regularization. In
addition to the results of \cite{Buchbinder:2016gmc}, we find all the
one-loop counterterms in the cut-off regularization scheme and show that
there is actually a counterterm proportional to $S_{SYM}$. In the
${\cal N}=(1,1)$ SYM theory, all the one-loop divergences are
canceled off shell.

\subsection{Two-point Green function of the gauge superfield}
\hspace*{\parindent}

\begin{figure}[h]
\begin{picture}(0,4)

\put(0.5,3.7){$(1)$}
\put(0.5,2.3){\includegraphics[scale=0.4]{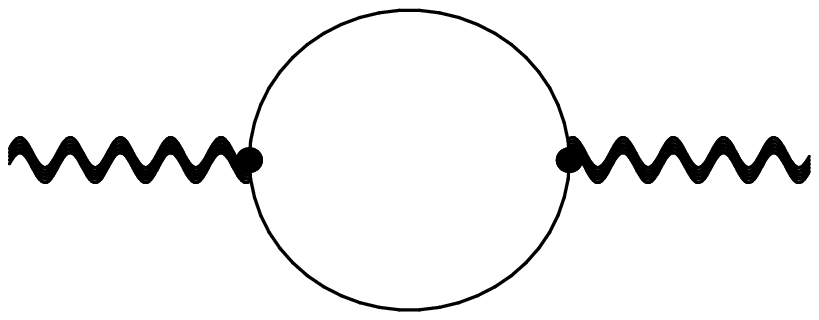}}
\put(4.5,3.7){$(2)$}
\put(4.5,2.3){\includegraphics[scale=0.4]{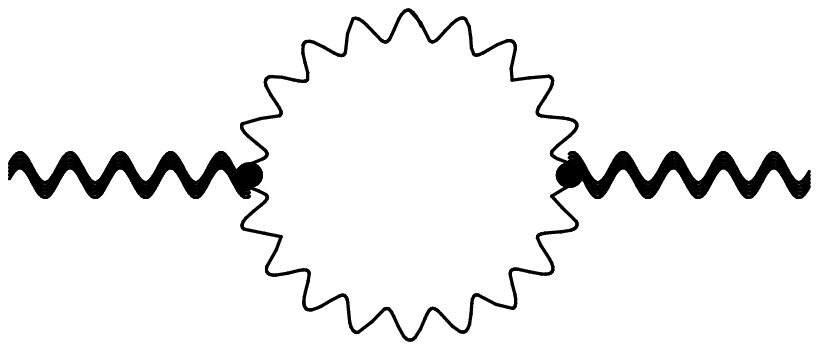}}
\put(8.5,3.7){$(3)$}
\put(8.5,2.3){\includegraphics[scale=0.4]{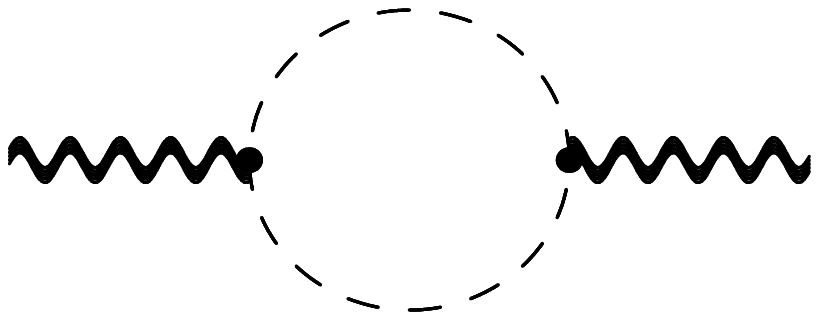}}
\put(12.5,3.7){$(4)$}
\put(12.5,2.3){\includegraphics[scale=0.4]{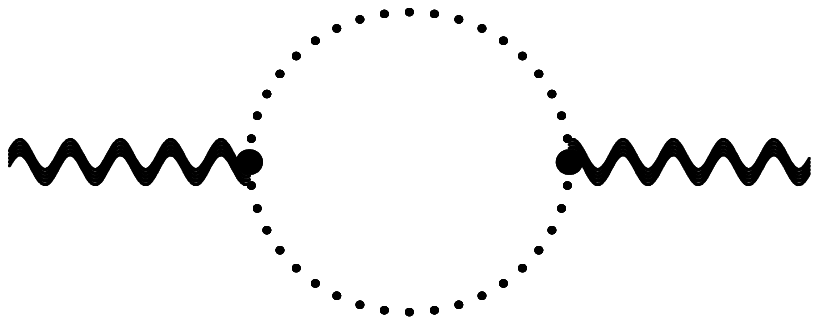}}

\put(2.8,1.5){$(5)$}
\put(3,0){\includegraphics[scale=0.4]{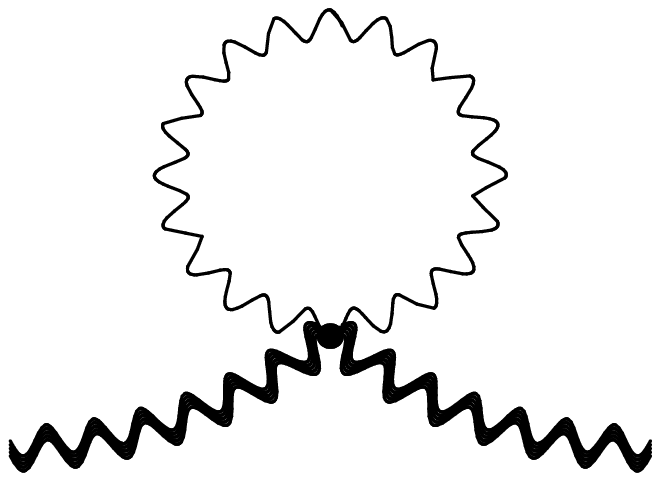}}
\put(6.8,1.5){$(6)$}
\put(7,0){\includegraphics[scale=0.4]{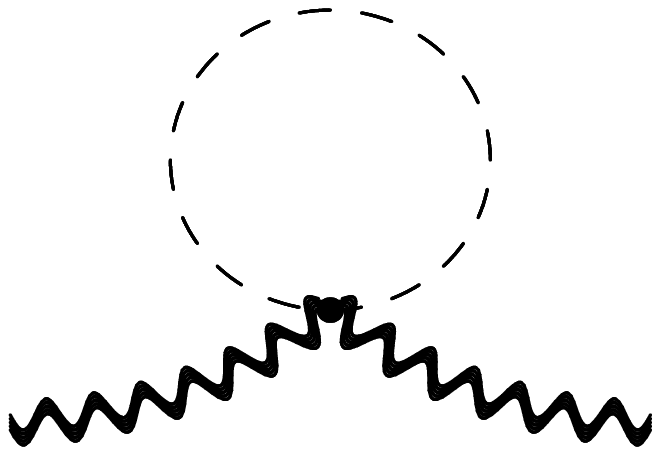}}
\put(10.8,1.5){$(7)$}
\put(11,0){\includegraphics[scale=0.4]{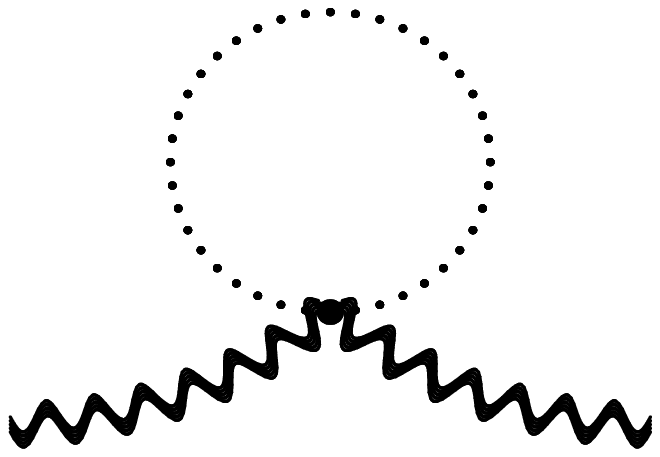}}

\end{picture}
\caption{One-loop contribution to the two-point Green
function of the background superfield in the abelian case.}
\label{Figure_Gauge_Harmonic_Diagrams}
\end{figure}

We start the computation of the one-loop divergences from the two-point Green function of the background gauge superfield.
In the considered approximation it is determined by the diagrams presented in Fig. \ref{Figure_Gauge_Harmonic_Diagrams}.
The bold external legs in these diagrams correspond to the background gauge superfield $\bm{V}^{++}$. Note that in the abelian case
the only non-trivial contribution comes from the diagram (1), while contributions of all other diagrams vanish. That is why we will start
our analysis by calculating the coefficient $c_1$ for the abelian theory (\ref{Action_Abelian}).

Using the standard rules of writing down the contributions of the harmonic supergraphs \cite{Galperin:2001uw},
%in the abelian case
we can present the contribution of the diagram (1) to the effective action of the abelian theory in the form

\begin{eqnarray}
&& \frac{i}{2} \int d\zeta^{(-4)}_1\, du_1\, d\zeta^{(-4)}_2\, du_2\, \bm{V}^{++}(z_1,u_1)
\bm{V}^{++}(z_2,u_2) \frac{1}{(u_2^+ u_1^+)^3} \frac{(D_1^+)^4 (D_2^+)^4}{\Box}\delta^{14}(z_1-z_2)\nonumber\\
&&\times \frac{1}{(u_1^+ u_2^+)^3} \frac{(D_1^+)^4 (D_2^+)^4}{\Box} \delta^{14}(z_1-z_2). \label{1zero}
\end{eqnarray}

\noindent
One of the two operator factors $(D_1^+)^4 (D_2^+)^4$ in \p{1zero} can be used to convert both integrations over $d\zeta^{(-4)}$ into those over $d^{14}z$,

\begin{equation}
- \frac{i}{2} \int d^{14}z_1\, du_1\, d^{14}z_2\, du_2\, \frac{\bm{V}^{++}(z_1,u_1) \bm{V}^{++}(z_2,u_2)}{(u_1^+ u_2^+)^6}
\frac{1}{\Box}\delta^{14}(z_1-z_2)\, \frac{(D_1^+)^4 (D_2^+)^4}{\Box} \delta^{14}(z_1-z_2).
\end{equation}

\noindent
After this, we should take into account that

\begin{equation}\label{Theta_Identity}
(D_1^+)^4 (D_2^+)^4 \delta^8(\theta_1-\theta_2) \, \delta^8(\theta_1-\theta_2) = (u_1^+ u_2^+)^4 \delta^8(\theta_1-\theta_2)
\end{equation}

\noindent
and perform one of the $\theta$-integrations with the help of the remaining Grassmann $\delta$-function $\delta^8(\theta_1-\theta_2)$. This gives

\begin{equation}
- \frac{i}{2} \int d^6x_1\, d^6x_2\, d^8\theta\, du_1\, du_2\, \bm{V}^{++}(x_1,\theta,u_1) \bm{V}^{++}(x_2,\theta,u_2)
\frac{1}{(u_1^+ u_2^+)^2} \frac{1}{\Box}\delta^{6}(x_1-x_2)\, \frac{1}{\Box} \delta^{6}(x_1-x_2).
\end{equation}

\noindent
Next, we rewrite this expression in the momentum space:
%The result can be presented in the form

\begin{equation}
-\frac{i}{2} \int \frac{d^6p}{(2\pi)^6} \int d^8\theta\, du_1\, du_2\, \bm{V}^{++}(p,\theta,u_1) \bm{V}^{++}(-p,\theta,u_2) \frac{1}{(u_1^+ u_2^+)^2}
\int \frac{d^6k}{(2\pi)^6} \frac{1}{k^2 (k+p)^2}.
\end{equation}

\noindent
Combining this expression with the tree-level result, the part of the effective action corresponding to the two-point function of the background
gauge superfield can be written as

\begin{eqnarray}
\Gamma^{(2)}_{\bm{V}^{++}} &=&  \int \frac{d^6p}{(2\pi)^6} \int d^8\theta\, du_1\, du_2\, \bm{V}^{++}(p,\theta,u_1) \bm{V}^{++}(-p,\theta,u_2) \frac{1}{(u_1^+ u_2^+)^2} \nonumber\\
&&\times \Big[\frac{1}{4f_0^2} - \frac{i}{2} \int \frac{d^6k}{(2\pi)^6} \frac{1}{k^2 (k+p)^2}\Big].\qquad
\end{eqnarray}

\noindent
{}From this expression we observe that the considered Green function is quadratically divergent.
To calculate the momentum integral, we resort to the standard trick of the Wick rotation to the Euclidean signature.
If $\Lambda$ is an UV cutoff, then, taking into account that the volume of the sphere $S^5$ is

\begin{equation}
\Omega_5 = \frac{2\pi^{(5+1)/2}}{\Gamma((5+1)/2)} = \frac{2\pi^3}{\Gamma(3)} = \pi^3,
\end{equation}

\noindent
the leading divergence of the considered Green function can be written in the form

\begin{equation}\label{Quadratic_Divergences}
\Gamma^{(2)}_{\infty,\bm{V}^{++}} =
\int \frac{d^6p}{(2\pi)^6} \int d^8\theta\, du_1\, du_2\, \bm{V}^{++}(p,\theta,u_1) \bm{V}^{++}(-p,\theta,-u_2) \frac{1}{(u_1^+ u_2^+)^2} \frac{\Lambda^2}{4(4\pi)^3}.
\end{equation}

\noindent
This expression is evidently gauge invariant, so there is no need to add any other quadratically divergent term with higher degrees of $\bm{V}^{++}$
to the one-loop effective action.
It is easy to see that in the coordinate representation it coincides, up to a numerical coefficient, with the classical action of the free gauge superfield

\begin{eqnarray}\label{Quadratic_Divergences1}
&& \int d^{14}z\, du_1\, du_2\, \bm{V}^{++}(z,u_1) \bm{V}^{++}(z,u_2) \frac{1}{(u_1^+ u_2^+)^2} \frac{\Lambda^2}{4(4\pi)^3} = \frac{f_0^2\Lambda^2}{(4\pi)^3} S_{U(1)}.
\end{eqnarray}

\noindent
This form of the quadratic divergence is in agreement with the results of Ref. \cite{Fradkin:1982kf}, where
the relation between leading divergences in various dimensions was analyzed. The $4D$ theory is renormalizable, so that the leading (logarithmic)
divergences in the gauge-field sector are proportional to $S_{U(1)}$. They are related with the leading (quadratic) divergences in the $6D$ theory,
which, thereby, should be also proportional to $S_{U(1)}$.

It is worth mentioning that  it is impossible to calculate quadratic divergences using the regularization scheme by dimensional reduction,
because the dimensional reduction technique does not see these divergences. This is the reason why for computing the quadratic divergences
one is led to use another regularization which do not break supersymmetries of the theory. Actually, the only regularization of this type is
the Slavnov higher-derivative regularization.
It was first proposed in \cite{Slavnov:1971aw,Slavnov:1972sq} for non-supersymmetric theories and was generalized to the supersymmetric case in
refs. \cite{Krivoshchekov:1978xg,West:1985jx}. For $4D$ ${\cal N}=2$ theories
it has been also worked out in the harmonic superspace approach \cite{Buchbinder:2015eva}. However, to generalize this result to the $6D$ case,
it is necessary to use regulators with higher degrees of covariant derivatives as compared with the $4D$ case. Now, this work is in progress.
Nevertheless, any version of the higher-derivative regularization will evidently produce Eq. (\ref{Quadratic_Divergences}) for the quadratic divergences.

If the theory is regularized through the dimensional reduction, it is possible to calculate only the logarithmic divergences.
In this case, using the standard Euclidean techniques, we obtain

\begin{equation}\label{Integral_Divergent_Part}
\int \frac{d^Dk}{(2\pi)^D} \frac{1}{k^2 (k+p)^2}  = -\frac{p^2}{3\varepsilon (4\pi)^3} + \mbox{finite terms},
\end{equation}

\noindent
where $\varepsilon \equiv 6-D$. Taking into account that $(p^2)_E = - (p^2)_M$, within the regularization by dimensional reduction the divergent part of the effective action
can be written as

\begin{equation}\label{Logarithmic_Divergence}
\int \frac{d^6p}{(2\pi)^6} \int d^8\theta\, du_1\, du_2\, \bm{V}^{++}(p,\theta,u_1) \bm{V}^{++}(-p,\theta,u_2) \frac{1}{(u_1^+ u_2^+)^2}
\frac{p^2}{6\varepsilon (4\pi)^3}.
\end{equation}

\noindent
It is known \cite{Bossard:2015dva} that the only gauge invariant expression of the considered dimension (in the abelian case) is given by

\begin{eqnarray}\label{Invariant}
&& \int d\zeta^{(-4)}\, du\, (\bm{F}^{++})^2 = \int d^{14}z\,du\, \bm{V}^{--} \Box \bm{V}^{++}\nonumber\\
&&\qquad\qquad\qquad\qquad\quad = \int d^{14}z\,du_1\,du_2\,\frac{1}{(u_1^+ u_2^+)^2} \bm{V}^{++}(z,u_1)\Box \bm{V}^{++}(z,u_2),\qquad
\end{eqnarray}

\noindent
where $\bm{F}^{++} \equiv (D^{+})^4 \bm{V}^{--}$ is a function of the background superfield $\bm{V}^{++}$. Comparing this expression
with Eq. (\ref{Logarithmic_Divergence}) we can present the logarithmical divergences in the gauge-field sector in the form

\begin{equation}\label{Gamma1}
\Gamma^{(1)}_{\infty,\ln} = -\frac{1}{6\varepsilon (4\pi)^3}\int d\zeta^{(-4)}\, du\, (\bm{F}^{++})^2 + \mbox{terms with hypermultiplets}
\end{equation}

\noindent
(the one-loop divergent contributions containing the hypermultiplet will be calculated below). Comparing Eq. (\ref{Gamma1}) with Eq. (\ref{Divergences}),
\footnote{As soon as we make calculations within the background field method, it is necessary to substitute $F^{++}$ by the similar expression $\bm{F}^{++}$
constructed from the background gauge superfield.} we conclude that the coefficient $c_1$ in Eq. (\ref{Divergences}) is

\begin{equation}
c_1 = -\frac{1}{6\varepsilon (4\pi)^3}.
\end{equation}

\noindent
This result agrees with the one obtained in \cite{Buchbinder:2016gmc} by the proper time technique. The coincidence of the results
derived by two different methods confirms the correctness of the calculations.

The calculation of the diagram (1) in the non-abelian case goes
along similar lines. The only novelty is the necessity to take into account the gauge group
indices and the factor

\begin{equation}
\mbox{tr}(T^A T^B) = T(R)\delta^{AB},
\end{equation}

\noindent
which comes from the generators appearing in the vertices. So in the non-abelian case the diagram (1) is represented by the expression

\begin{equation}\label{D1_Non_Abelian}
-\frac{i}{2} T(R) \int \frac{d^6p}{(2\pi)^6} \int d^8\theta\, du_1\, du_2\, \bm{V}^{++A}(p,\theta,u_1) \bm{V}^{++A}(-p,\theta,u_2)
\frac{1}{(u_1^+ u_2^+)^2} \int \frac{d^6k}{(2\pi)^6} \frac{1}{k^2 (k+p)^2}.
\end{equation}

The contributions of the other diagrams depicted in Fig. \ref{Figure_Gauge_Harmonic_Diagrams} are calculated in Appendix A2.
%\ref{Appendix_Two_Point}.
There we demonstrate that the sum of all diagrams containing the loop of the quantum gauge superfield (i.e. (2) and (5)) vanishes and the net result
comes solely from the ghost contributions. Adding the latter to Eq. (\ref{D1_Non_Abelian}), we obtain the total two-point function of the gauge superfield in the form

\begin{eqnarray}\label{Quadratic}
&& \frac{i}{2} \Big[C_2 - T(R)\Big] \int \frac{d^6p}{(2\pi)^6} \int d^8\theta\,
du_1\, du_2\, \bm{V}^{++A}(p,\theta,u_1) \bm{V}^{++A}(-p,\theta,u_2)\frac{1}{(u_1^+ u_2^+)^2}\qquad \nonumber\\
&& \times  \int \frac{d^6k}{(2\pi)^6} \frac{1}{k^2 (k+p)^2}.
\end{eqnarray}

\noindent
The divergent part of this expression is calculated in the precisely same way as in the abelian case. In particular, the leading quadratic divergence can be written as

\begin{equation}\label{Logarithm}
-\Big[C_2-T(R)\Big]\int d^{14}z\, du_1\, du_2\, \bm{V}^{++ A}(z,u_1) \bm{V}^{++ A}(z,u_2) \frac{1}{(u_1^+ u_2^+)^2} \frac{\Lambda^2}{4(4\pi)^3}.
\end{equation}

\noindent
In the case of using the regularization by dimensional reduction the (logarithmical) divergence is parametrized by the expression

\begin{equation}\label{D1}
\Big[C_2 - T(R)\Big] \int d^{14}z\,du_1\, du_2\, \bm{V}^{++ A}(z,u_1) \Box \bm{V}^{++ A}(z,u_2) \frac{1}{(u_1^+ u_2^+)^2}
\frac{1}{6\varepsilon (4\pi)^3}.
\end{equation}

\noindent
It is worth to point out an essential difference of these results from their abelian counterparts. Actually, in the non-abelian case both $\bm{V}^{--}$ and $\bm{F}^{++}$
are non-linear functions of $\bm{V}^{++}$. This implies that there should be also divergent contributions proportional to higher degrees of $\bm{V}^{++}$.
However, they can be easily restored by taking into account the gauge invariance of the action $S_{\mbox{\scriptsize SYM}}$ and the expression $\mbox{tr}(\bm{F^{++}})^2$.
Comparing quadratic terms in these expressions with (\ref{Quadratic}) and (\ref{Logarithm}), respectively, we find that the leading quadratic divergence is

\begin{equation}
-\Big[C_2-T(R)\Big]\frac{f_0^2 \Lambda^2}{(4\pi)^3}\, S_{SYM}[\bm{V}^{++}],
\end{equation}

\noindent
while the logarithmic divergence obtained with making use of the dimensional reduction has the form

\begin{equation}
\frac{C_2-T(R)}{3\varepsilon (4\pi)^3} \mbox{tr} \int d\zeta^{(-4)}\, du\, (\bm{F}^{++})^2\,,
\end{equation}

\noindent
that coincides with the result obtained in \cite{Buchbinder:2016url}. We see that both these expressions vanish
in the case of ${\cal N}=(1,1)$ SYM theory, for which $T(R)=C_2$.

Note that the vanishing of quadratic divergences is quite expectable for ${\cal N}=(1,1)$ theory.
Actually, the quadratic divergences can appear only in the gauge-multiplet sector, while the formula for the degree of divergence (\ref{Divergence_Degree})
tells us that in the sector involving hypermultiplets only the logarithmically divergent terms can appear. On the other hand,
the hidden ${\cal N}=(0,1)$ symmetry would imply the appearance of quadratic divergences in the hypermultiplet sector, once they would be present in the gauge-multiplet sector.
Since no such divergences are possible in the hypermultiplet sector, the quadratic divergences cannot appear in ${\cal N}=(1,1)$ theory at all.

\subsection{Two-point Green function of the hypermultiplet}
\hspace*{\parindent}

\begin{figure}[h]
\begin{picture}(0,2)
\put(6,0){\includegraphics[scale=0.4]{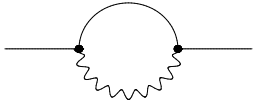}}
\end{picture}
\caption{The diagram contributing to the two-point Green function of the hypermultiplet}\label{Figure_Hypermultiplet}
\end{figure}

Let us now calculate the two-point Green function of the matter hypermultiplet superfields. In the one-loop approximation
it is determined by the diagram depicted in Fig. \ref{Figure_Hypermultiplet}, for which $N_q=2$. Then, according to Eq. (\ref{Divergence_Degree}),
it is logarithmically divergent.

\noindent
Note that the result for this diagram is gauge-dependent due to the presence of the gauge-superfield propagator. However, in this paper we do calculations
only in the minimal gauge corresponding to the choice $\xi_0=1$. Then the expression constructed according to the Feynman rules has the form

\begin{eqnarray}
&& -2if_0^2 \int d\zeta^{(-4)}_1\, du_1\, d\zeta^{(-4)}_2\, du_2\, \widetilde q^+(z_1,u_1)^i (T^A T^A)_i{}^j q^+(z_2,u_2)_j \frac{1}{(u_1^+ u_2^+)^3} \qquad\nonumber\\
&& \times \frac{(D_1^+)^4 (D_2^+)^4}{\Box} \delta^{14}(z_1-z_2) \frac{1}{\Box} (D_1^+)^4 \delta^{(2,-2)}(u_2,u_1) \delta^{14}(z_1-z_2).\label{2zero}
\end{eqnarray}

\noindent
As in the case considered in the previous Subsection, the product of derivatives $(D_1^+)^4 (D_2^+)^4$ present in \p{2zero} makes
it possible to restore two integrations over $d^{14}z$,

\begin{eqnarray}
&& -2if_0^2 \int d^{14}z_1\, du_1\, d^{14}z_2\, du_2\, \widetilde q^+(z_1,u_1)^i C(R)_i{}^j q^+(z_2,u_2)_j \frac{1}{(u_1^+ u_2^+)^3}\,
\frac{1}{\Box} \delta^{14}(z_1-z_2)\qquad\nonumber\\
&& \times \frac{1}{\Box} (D_1^+)^4 \delta^{(2,-2)}(u_2,u_1) \delta^{14}(z_1-z_2) = 0.\label{3zero}
\end{eqnarray}

\noindent
The vanishing of this expression follows from the property that the product of Grassmann $\delta$-functions at coincident points
does not vanish only provided there are at least eight spinor covariant derivatives acting on one of them, which is not the case for \p{3zero}.
Therefore, in the minimal gauge the hypermultiplets are not renormalized in the one-loop approximation. Obviously, this result is valid in both non-abelian
and abelian cases.

\subsection{Three-point gauge-hypermultiplet Green function}
\hspace*{\parindent}

\begin{figure}[h]
\begin{picture}(0,3)
\put(3.9,0.){\includegraphics[scale=0.18]{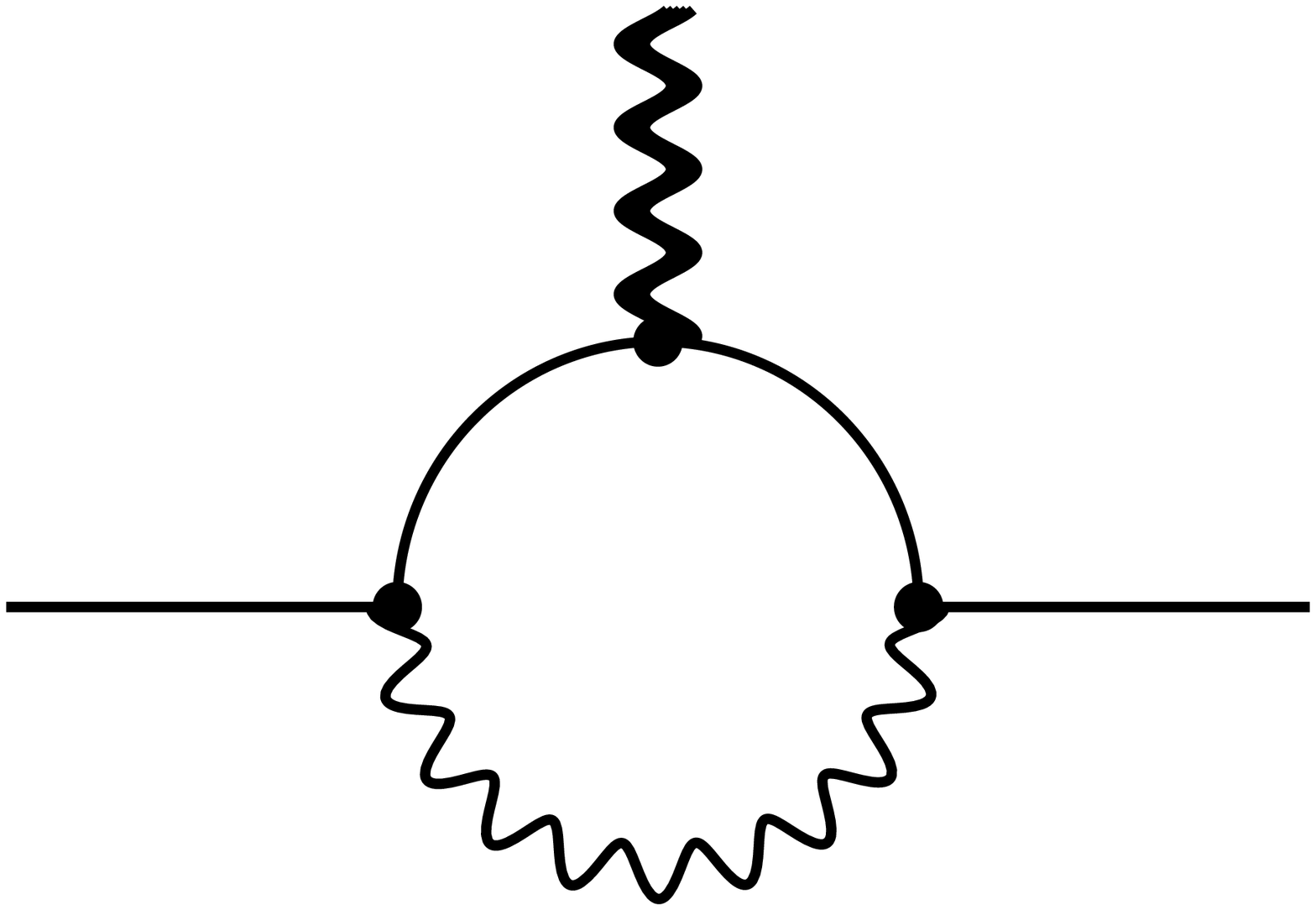}}
\put(3.9,2){$(1)$}
\put(8.9,0.15){\includegraphics[scale=0.18]{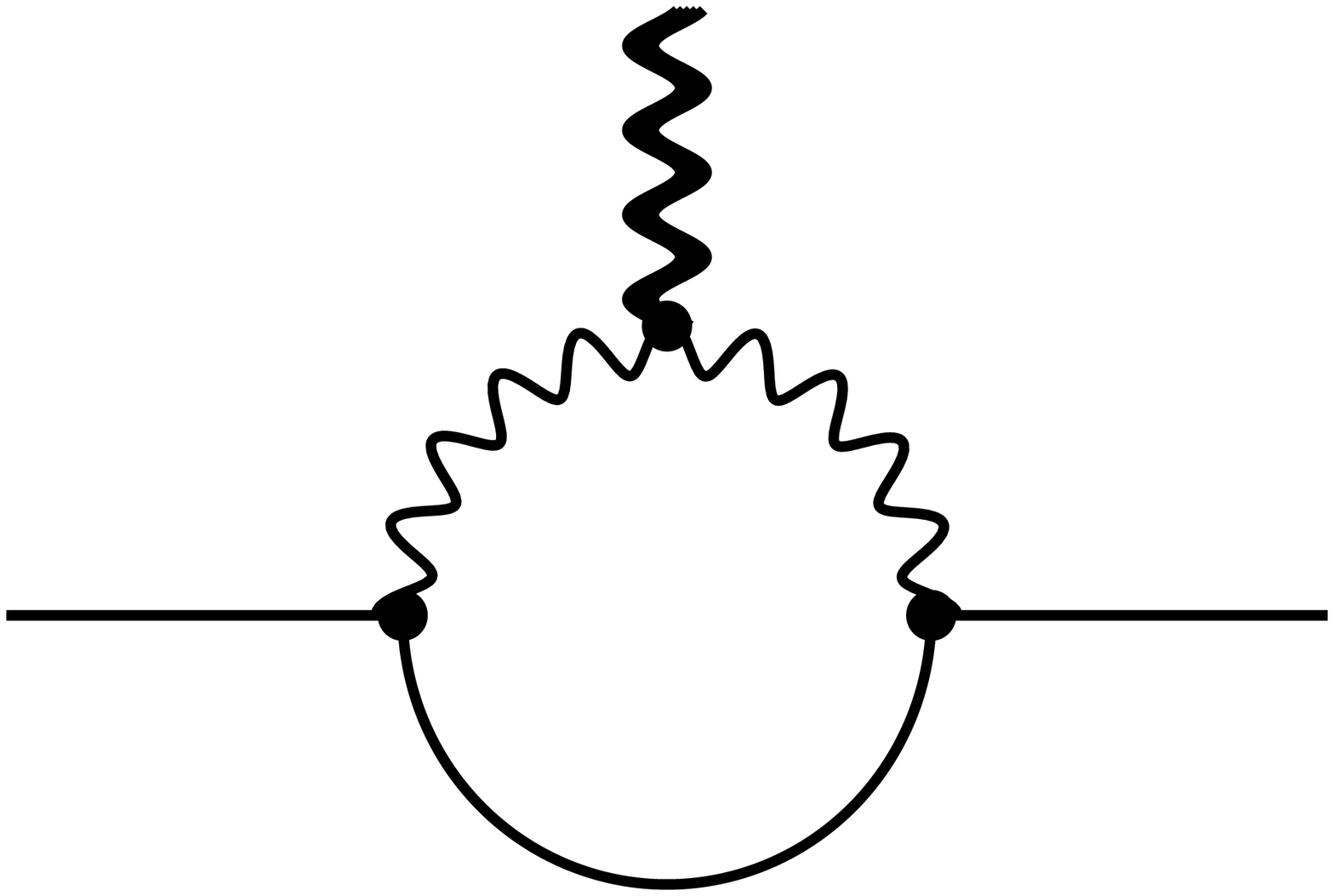}}
\put(8.9,2){$(2)$}
\end{picture}
\caption{The diagrams which determine the three-point gauge-hypermultiplet function. In the abelian case only the diagram (1) is non-vanishing.}\label{Figure_One-Loop_Vertex}
\end{figure}

In order to determine the coefficient $c_2$ in Eq. (\ref{Divergences}),
it suffices to consider the three-point gauge-hypermultiplet Green function, which in the one-loop approximation
is represented by the diagrams depicted in Fig. \ref{Figure_One-Loop_Vertex}. Evidently, in the abelian case only the diagram (1) remains.
This is why we will start with studying the one-loop divergence in the abelian case. Again, we will calculate the corresponding contribution
to the effective action in the minimal gauge $\xi_0=1$. For the abelian theory (\ref{Action_Abelian}) it has the form

\begin{eqnarray}
&&\hspace*{-5mm} - 2f_0^2 \int d\zeta^{(-4)}_1\,du_1\,d\zeta^{(-4)}_2\,du_2\,d\zeta^{(-4)}_3\,du_3\,\widetilde q^+(z_1,u_1)
q^+(z_3,u_3)\bm{V}^{++}(z_2,u_2) \frac{(D_1^+)^4}{\Box} \delta^{(2,-2)}(u_3,u_1)\nonumber\\
&&\hspace*{-5mm}\times \delta^{14}(z_1-z_3)\, \frac{1}{(u_1^+ u_2^+)^3}\frac{(D_1^+)^4 (D_2^+)^4}{\Box} \delta^{14}(z_1-z_2)
\, \frac{1}{(u_2^+ u_3^+)^3}\frac{(D_2^+)^4 (D_3^+)^4}{\Box} \delta^{14}(z_2-z_3).
\end{eqnarray}

\noindent
The calculation of this diagram is described in Appendix B1.
%\ref{Appendix_3Point_Abelian}.
In the momentum representation (in the Minkowski space) the result can be presented in the form

\begin{eqnarray}
&& -2f_0^2 \int \frac{d^{6}p}{(2\pi)^6}\,\frac{d^{6}q}{(2\pi)^6}\,d^8\theta\,du_1\,du_2\,\widetilde q^+(q+p,\theta,u_1) q^+(-q,\theta,u_1)\bm{V}^{++}(-p,\theta,u_2)
\frac{1}{(u_1^+ u_2^+)^2}\nonumber\\
&& \times \int \frac{d^{6}k}{(2\pi)^6}\, \frac{1}{k^2 (q+k)^2 (q+k+p)^2}.
\end{eqnarray}

\noindent
To find a divergent part of this expression, we perform the Wick rotation in the integral over the loop momentum $k$ and rewrite it in the Euclidean space as

\begin{equation}
2i f_0^2 \int \frac{d^6k}{(2\pi)^6}\frac{1}{k^2 (k+q)^2 (k+q+p)^2}.
\end{equation}

\noindent Evidently, this expression is logarithmically divergent.
So far, it is not well-defined, because we have not yet specified
the regularization. Within the dimensional reduction method
\cite{Siegel:1979wq}, it is necessary to substitute $D=6$ by
$D=6-\varepsilon$ with $\varepsilon\ne 0$. Then, using the standard
technique, we find

\begin{equation}\label{Triple_Integral}
\int \frac{d^Dk}{(2\pi)^6} \frac{1}{k^2 (k+q)^2 (k+q+p)^2} = \frac{1}{\varepsilon (4\pi)^3} + \mbox{finite terms}.
\end{equation}

\noindent
Thus, in the configuration space the divergent part of the considered contribution to the effective action is given by

\begin{eqnarray}
&& 2i f_0^2 \frac{1}{\varepsilon(4\pi)^3}\int d^{14}z\, \int du_1\,du_2\, \widetilde q^+(z,u_1)\, \bm{V}^{++}(z,u_2)\, q^+(z,u_1) \frac{1}{(u_1^+ u_2^+)^2}\nonumber\\
&& = 2i f_0^2 \frac{1}{\varepsilon(4\pi)^3}\int d^{14}z\, \int du\, \widetilde q^+(z,u)\, \bm{V}^{--}(z,u)\, q^+(z,u).\label{5zero}
\end{eqnarray}

\noindent
%{}From Eqs. (\ref{Integrations}) and (\ref{Measure}) we see that
Rewriting

\begin{equation}
\int d^{14}z = \int d\zeta^{(-4)}\,(D^+)^4\,,
\end{equation}

\noindent
and taking into account the analyticity of the superfields $\widetilde q^+$ and $q^+$, the expression \p{5zero} can be written in terms
of $\bm{F}^{++} = (D^+)^4 \bm{V}^{--}$ as

\begin{equation}
2i f_0^2 \frac{1}{\varepsilon(4\pi)^3}\int d\zeta^{(-4)}\, du\, \widetilde q^+ (D^+)^4 \bm{V}^{--} q^+ = 2i f_0^2
\frac{1}{\varepsilon(4\pi)^3}\int d\zeta^{(-4)}\, du\, \widetilde q^+ \bm{F}^{++} q^+.
\end{equation}

\noindent
Comparing this expression with Eq. (\ref{Divergences}), we find that

\begin{equation}
c_2 = \frac{2f_0^2}{\varepsilon(4\pi)^3}.
\end{equation}

\noindent
This result coincides with the one obtained in \cite{Buchbinder:2016gmc} by the proper time technique.

Next, let us consider the non-abelian case. In this case both diagrams in
Fig. \ref{Figure_One-Loop_Vertex} contribute to the Green
function. The diagram (1) is calculated in a close analogy with the abelian case.
It can be represented as

\begin{eqnarray}
&& - 2f_0^2 \int d\zeta^{(-4)}_1\,du_1\,d\zeta^{(-4)}_2\,du_2\,d\zeta^{(-4)}_3\,du_3\,
\widetilde q^+(z_1,u_1)^i \big(T^A\bm{V}^{++}(z_2,u_2)T^A\big)_i{}^j q^+(z_3,u_3)_j \nonumber\\
&&\times \frac{(D_1^+)^4}{\Box} \delta^{(2,-2)}(u_3,u_1) \delta^{14}(z_1-z_3)\,
\frac{1}{(u_1^+ u_2^+)^3}\frac{(D_1^+)^4 (D_2^+)^4}{\Box} \delta^{14}(z_1-z_2)\, \frac{1}{(u_2^+ u_3^+)^3}\qquad\nonumber\\
&& \times \frac{(D_2^+)^4 (D_3^+)^4}{\Box} \delta^{14}(z_2-z_3).
\end{eqnarray}

\noindent
Repeating the calculation steps described above and also taking into account that

\begin{eqnarray}
&& T^A \bm{V}^{++} T^A = \bm{V}^{++B} T^A T^B T^A =  \bm{V}^{++B} \Big(T^A T^A T^B + T^A [T^B, T^A]\Big)\qquad\nonumber\\
&&  = \bm{V}^{++B} \Big[C(R) T^B - \frac{1}{2} C_2 T^B\Big] = \Big[C(R)-\frac{1}{2}C_2\Big] \bm{V}^{++},
\end{eqnarray}

\noindent
we obtain the expression for the diagram (1) in the form

\begin{equation}\label{Diagram1}
2i f_0^2 \frac{1}{\varepsilon(4\pi)^3}\int d\zeta^{(-4)}\, du\, (\widetilde q^+)^i
\Big[C(R)_i{}^k-\frac{1}{2} C_2\delta_i^k\Big] (\bm{F}_{\mbox{\scriptsize linear}}^{++})_k{}^j (q^+)_j,
\end{equation}

\noindent
where
%the subscript "linear"\, denotes the linear  of the corresponding expression,

\begin{equation}
\bm{F}_{\mbox{\scriptsize linear}}^{++} \equiv (D^{+})^4 \bm{V}_{\mbox{\scriptsize linear}}^{--} \equiv (D^+)^4 \int du_1 \frac{\bm{V}^{++}(z,u_1)}{(u^+ u_1^+)^2}.
\end{equation}
The nonlinear terms will be restored below by the gauge-invariance reasoning.

Let us turn to calculating the second diagram in Fig. \ref{Figure_One-Loop_Vertex}. The details of this calculation are described
in Appendix B.
%\ref{Appendix_3Point_Non-Abelian}.
The expression for the diagram (2) in Fig. \ref{Figure_One-Loop_Vertex} obtained there is as follows

\begin{eqnarray}
&& f_0^2 C_2 \int \frac{d^6p}{(2\pi)^6} \frac{d^6q}{(2\pi)^6} \frac{d^6k}{(2\pi)^6} d^8\theta\,
\frac{1}{k^2 (k+p)^2 (k+p+q)^2} \int du\, \widetilde q^+(q+p,\theta,u)^i \bm{V}_{\mbox{\scriptsize linear}}^{--}(-p,\theta,u)_i{}^j \nonumber\\
&& \times q^+(-q,\theta,u)_j.
\end{eqnarray}

\noindent
The integral over the loop momentum $k$ is calculated using the Wick rotation. In the case of the regularization by dimensional reduction
it can be found based on the result (\ref{Triple_Integral}). Then, the divergent part reads

\begin{eqnarray}
&&-\frac{i}{\varepsilon (4\pi)^3} f_0^2 C_2 \int d^{14}z\,du (\widetilde q^+)^i (\bm{V}_{\mbox{\scriptsize linear}}^{--})_i{}^j (q^+)_j \nonumber \\
&& =\, -\frac{i}{\varepsilon (4\pi)^3} f_0^2 C_2 \int d\zeta^{(-4)}\,du (\widetilde q^+)^i (\bm{F}_{\mbox{\scriptsize linear}}^{++})_i{}^j (q^+)_j.
\end{eqnarray}

Adding this contribution to Eq. (\ref{Diagram1}), we obtain the total result for the diagrams presented in Fig. \ref{Figure_One-Loop_Vertex},

\begin{equation}\label{Vertex_Result}
2i f_0^2 \frac{1}{\varepsilon(4\pi)^3}\int d\zeta^{(-4)}\, du\, (\widetilde q^+)^i \Big[C(R)_i{}^k- C_2\delta_i^k\Big] (\bm{F}_{\mbox{\scriptsize linear}}^{++})_k{}^j (q^+)_j.
\end{equation}

\noindent
It is linear in $\bm{V}^{++}$ by construction. However, we know that the result for the one-loop divergences should be gauge invariant.
The only possible gauge invariant expression yielding Eq. (\ref{Vertex_Result}) in the linearization limit is as follows

\begin{equation}
2i f_0^2 \frac{1}{\varepsilon(4\pi)^3}\int d\zeta^{(-4)}\, du\, (\widetilde q^+)^i \Big[C(R)_i{}^k- C_2\delta_i^k\Big] (\bm{F}^{++})_k{}^j (q^+)_j.\label{6zero}
\end{equation}

\noindent
It agrees with the result obtained in \cite{Buchbinder:2016url} by the proper time technique. Choosing the representation $R$ to be irreducible,
we obtain the following value for the coefficient $c_2$:

\begin{equation}
c_2 = 2 f_0^2\, \frac{C(R)-C_2}{(4\pi)^3 \varepsilon}\,.
\end{equation}

\noindent
{}We see that the corresponding divergence vanishes for ${\cal N}=(1,1)$ SYM theory, when $R$ is adjoint representation.

\subsection{Total one-loop divergences of the theory}
\hspace*{\parindent}

Let us summarize the results obtained in the previous section and write down the total divergent part of the effective action
for $6D\,,$ ${\cal N}=(1,0)$ SYM theory. If this theory is regularized through the dimensional reduction, then

\begin{eqnarray}\label{Gamma_DRED}
&& (\Gamma^{(1)}_{\infty})_{\mbox{\scriptsize DRED}} = \frac{C_2-T(R)}{3\varepsilon (4\pi)^3} \mbox{tr} \int d\zeta^{(-4)}\, du\, (\bm{F}^{++})^2\nonumber\\
&&\qquad\qquad\qquad -\,  2i f_0^2 \frac{1}{\varepsilon(4\pi)^3}\int d\zeta^{(-4)}\, du\, \widetilde q^+ \big[C_2-C(R)\big]\bm{F}^{++} q^+.\qquad
\end{eqnarray}

\noindent
However, as we have already mentioned, within the dimensional reduction technique it is impossible to catch quadratic divergencies.
They can be obtained, for example, using the momentum cut-off regularization. Then the result for the one-loop divergences can be written as

\begin{eqnarray}\label{Gamma_CutOff}
&& (\Gamma^{(1)}_{\infty})_{\mbox{\scriptsize UV cut-off}} = -\big[C_2-T(R)\big]\frac{f_0^2 \Lambda^2}{(4\pi)^3}\, S_{SYM}[\bm{V}^{++}]
+ \ln\Lambda\Big[\frac{C_2-T(R)}{3 (4\pi)^3} \mbox{tr}\int d\zeta^{(-4)}\, du\, \qquad\nonumber\\
&& \times (\bm{F}^{++})^2 - 2i f_0^2 \frac{1}{(4\pi)^3}\int d\zeta^{(-4)}\, du\, \widetilde q^+ \big[C_2-C(R)\big] \bm{F}^{++} q^+\Big],
\end{eqnarray}

\noindent
where $\Lambda$ denotes the ultraviolet cut-off. The first term corresponds to the quadratic divergences, while the remaining two terms
parametrize the logarithmic divergences.

In the abelian case the corresponding expressions take the form

\begin{eqnarray}\label{Gamma_DRED_Abelian}
&& (\Gamma^{(1)}_{\infty})_{\mbox{\scriptsize DRED}} = -\frac{1}{6\varepsilon (4\pi)^3}\int d\zeta^{(-4)}\,
du\, (\bm{F}^{++})^2 +  2i f_0^2 \frac{1}{\varepsilon(4\pi)^3}\int d\zeta^{(-4)}\, du\, \widetilde q^+ \bm{F}^{++} q^+;\nonumber\\
\label{Gamma_CutOff_Abelian}
&& (\Gamma^{(1)}_{\infty})_{\mbox{\scriptsize UV cut-off}} = \int d^{14}z\, du_1\, du_2\, \bm{V}^{++}(z,u_1) \bm{V}^{++}(z,u_2) \frac{1}{(u_1^+ u_2^+)^2}
\frac{\Lambda^2}{4(4\pi)^3}\nonumber\\
&& + \ln\Lambda\Big[-\frac{1}{6 (4\pi)^3}\int d\zeta^{(-4)}\, du\, (\bm{F}^{++})^2 +  2i f_0^2 \frac{1}{(4\pi)^3}\int d\zeta^{(-4)}\,
du\, \widetilde q^+ \bm{F}^{++} q^+\Big].
\end{eqnarray}

\section{Summary}
\hspace*{\parindent}\label{Section_Summary}

In this paper we have studied the quantum aspects of generic
supersymmetric $6D, \,{\cal N}=(1,0)$ gauge theory of interacting six-dimensional
gauge multiplet minimally coupled to hypermultiplet. The theory is
formulated in ${\cal N}=(1,0)$ harmonic
superspace that allows one to preserve the manifest ${\cal N}=(1,0)$
supersymmetry at all steps of consideration. Also we used the superfield background field method
that secures, besides manifest
supersymmetry, the manifest classical gauge invariance of quantum
theory. The $6D, \,{\cal N}=(1,0)$ harmonic supergraph technique was
developed to study the off-shell effective action depending on
the gauge and hypermultiplet superfields and it was applied for
calculating the one-loop divergences. We have considered both
abelian and non-abelian ${\cal N}=(1,0)$ models and also $6D, {\cal
N}=(1,1)$ SYM theory as a particular case of the general system.

We investigated the divergent part of the one-loop effective action
corresponding to the two- and three-point functions for $6D,$ ${\cal
N}=(1,0)$ SYM theory interacting with hypermultiplets. Using the
supergraph techniques in harmonic superspace we calculated the
two-point Green functions of both the gauge superfield and the
hypermultiplet. Also we found the three-point mixed gauge-hypermultiplet
Green function. The results for these Green functions allowed us to
restore the total gauge invariant result for the off-shell
one-loop divergences in the theory under consideration. The calculations were
performed for both abelian and non-abelian models.

In the non-abelian case it was demonstrated  that the divergences reveal a
generic structure, first found in \cite{Buchbinder:2016url} on the
basis of the operator proper-time technique. Namely, all of these divergences are proportional
to the difference of Casimir operators for the adjoint representation and
representation $R$ to which the hypermultiplet belongs. This leads us to
conclude that the $6D, {\cal N}=(1,1)$ SYM theory is completely
off-shell finite in the one-loop approximation. The results for
abelian theory are also consistent with the earlier calculations in
\cite{Buchbinder:2016gmc} where they were done by another
method. It is worth pointing out that the calculations in terms of
supergraphs are more transparent and simpler then those within
the operator proper-time techniques as performed in \cite{Buchbinder:2016gmc,Buchbinder:2016url}. It should be mentioned that, besides the
logarithmic divergences calculated earlier in \cite{Buchbinder:2016gmc,Buchbinder:2016url}, in the present paper we calculated the power
divergences which also have an interesting structure and vanish for ${\cal N}=(1,1)$ SYM theory as well.

The absence of off-shell one-loop divergences in
$6D, \,{\cal N}=(1,1)$ SYM theory raises a question concerning the
off-shell structure of higher-loop divergences in this theory. We
would like to recall that the divergences in the hypermultiplet
sector have never been considered even on shell. The supergraph
technique, formulated in this paper, provides a natural ground for higher-loop
calculations. Note that, according to standard renormalization
procedure, the calculations of the Feynman diagrams for the higher-loop
divergences implicate that the divergences of all sub-diagrams have
been already eliminated with the help of the lower-loop counterterms. Since
$6D, {\cal N}=(1,1)$ SYM theory is off-shell one-loop finite, the
analysis of two-loop divergences is simplified because the one-loop
divergences are absent and there is no need to renormalize the one-loop
subgraphs. We plan to study the complete off-shell structure of the
two-loop divergences of this theory in a forthcoming work.

\section*{Acknowledgements} \hspace*{\parindent}
We are very grateful to I.B.Samsonov for valuable discussions. This
work was supported by the grant of Russian Scientific Foundation,
project No. 16-12-10306.

\renewcommand\theequation{A.\arabic{equation}} \setcounter{equation}0
\section*{Appendix A\quad}

\section*{Two-point Green function of the background gauge superfield}
\label{Appendix_Two_Point}

\subsection*{A1. Contribution of the Faddeev--Popov and Nielsen--Kallosh ghosts}
\hspace{\parindent}

Let us calculate the diagrams (3) and (6) in Fig. \ref{Figure_Gauge_Harmonic_Diagrams} that correspond to the contribution of the Faddeev--Popov ghosts.
The diagram (3) is given by the expression

\begin{eqnarray}
&&\hspace*{-8mm} i \int d\zeta^{(-4)}_1\, d\zeta^{(-4)}_2\, du_1\, du_2\, f^{ABC} f^{CDA} \bm{V}^{++ B}(z_1,u_1) \bm{V}^{++ D}(z_2,u_2)\nonumber\\
&&\hspace*{-8mm} \times \Bigg[D_2^{++}\Big[\frac{(D_1^+)^4 (D_2^+)^4}{\Box} \delta^{14}(z_1-z_2) \frac{(u_1^- u_2^-)}{(u_1^+ u_2^+)^3}\Big] D_1^{++}
\Big[\frac{(D_1^+)^4 (D_2^+)^4}{\Box} \delta^{14}(z_1-z_2) \frac{(u_1^- u_2^-)}{(u_1^+ u_2^+)^3}\Big]\nonumber\\
&&\hspace*{-8mm} - \Big[\frac{(D_1^+)^4 (D_2^+)^4}{\Box} \delta^{14}(z_1-z_2) \frac{(u_1^- u_2^-)}{(u_1^+ u_2^+)^3}\Big]
D_1^{++} D_2^{++} \Big[\frac{(D_1^+)^4 (D_2^+)^4}{\Box} \delta^{14}(z_1-z_2) \frac{(u_1^- u_2^-)}{(u_1^+ u_2^+)^3}\Big] \Bigg].
\end{eqnarray}

\noindent
Using the relations (\ref{Integrations}), we rewrite it as

\begin{eqnarray}
&&\hspace*{-8mm}  - i C_2 \int d^{14}z_1\, d^{14}z_2\, du_1\, du_2\, \bm{V}^{++ A}(z_1,u_1) \bm{V}^{++ A}(z_2,u_2)\nonumber\\
&&\hspace*{-8mm} \times \Bigg[D_2^{++}\Big[\frac{1}{\Box} \delta^{14}(z_1-z_2) \frac{(u_1^- u_2^-)}{(u_1^+ u_2^+)^3}\Big] D_1^{++}
\Big[\frac{(D_1^+)^4 (D_2^+)^4}{\Box} \delta^{14}(z_1-z_2) \frac{(u_1^- u_2^-)}{(u_1^+ u_2^+)^3}\Big]\nonumber\\
&&\hspace*{-8mm} - \Big[\frac{1}{\Box} \delta^{14}(z_1-z_2) \frac{(u_1^- u_2^-)}{(u_1^+ u_2^+)^3}\Big] D_1^{++} D_2^{++}
\Big[\frac{(D_1^+)^4 (D_2^+)^4}{\Box} \delta^{14}(z_1-z_2) \frac{(u_1^- u_2^-)}{(u_1^+ u_2^+)^3}\Big] \Bigg].
\end{eqnarray}

\noindent
In order to get rid of the integrals over the anticommuting variables, we first apply the identity (\ref{Theta_Identity}) and,
then, calculate one of the anticommuting integral by making use of the $\delta$-function $\delta^8(\theta_1-\theta_2)$. This gives

\begin{eqnarray}
&& - i C_2 \int d^{6}x_1\, d^{6}x_2\, d^8\theta\, du_1\, du_2\, \bm{V}^{++ A}(x_1,\theta,u_1) \bm{V}^{++ A}(x_2,\theta,u_2) (u_1^+ u_2^+)^4 \nonumber\\
&& \times \Bigg[\frac{1}{\Box} \delta^{6}(x_1-x_2) D_2^{++}\Big[\frac{(u_1^- u_2^-)}{(u_1^+ u_2^+)^3}\Big] \, \frac{1}{\Box} \delta^{6}(x_1-x_2)
D_1^{++}\Big[\frac{(u_1^- u_2^-)}{(u_1^+ u_2^+)^3}\Big]\nonumber\\
&& - \frac{1}{\Box} \delta^{6}(x_1-x_2) \frac{(u_1^- u_2^-)}{(u_1^+ u_2^+)^3}\, \frac{1}{\Box} \delta^{6}(x_1-x_2) D_1^{++} D_2^{++}
\Big[\frac{(u_1^- u_2^-)}{(u_1^+ u_2^+)^3}\Big] \Bigg].
\end{eqnarray}

\noindent
At the next step, we rewrite this expression in the momentum representation and compute the harmonic derivatives,

\begin{eqnarray}
&&  - i C_2 \int \frac{d^6p}{(2\pi)^6}\,  d^8\theta\, du_1\, du_2\, \bm{V}^{++ A}(-p,\theta,u_1) \bm{V}^{++ A}(p,\theta,u_2)  \nonumber\\
&& \times \int \frac{d^6k}{(2\pi)^6}\, \frac{1}{k^2(k+p)^2} \Big[ -\frac{(u_2^+ u_1^-)(u_1^+ u_2^-)}{(u_1^+ u_2^+)^2} - \frac{(u_1^- u_2^-)}{(u_1^+ u_2^+)}
\Big].
\end{eqnarray}

\noindent
To simplify it, we use the identity

\begin{equation}
(u_2^+ u_1^-)(u_1^+ u_2^-) = 1 - (u_1^- u_2^-) (u_1^+ u_2^+).
\end{equation}

\noindent
Then, for the diagram (3) we obtain

\begin{equation}
i C_2 \int \frac{d^6p}{(2\pi)^6}\,  d^8\theta\, du_1\, du_2\, \bm{V}^{++ A}(-p,\theta,u_1) \bm{V}^{++ A}(p,\theta,u_2) \frac{1}{(u_1^+u_2^+)^2}
\int \frac{d^6k}{(2\pi)^6}\, \frac{1}{k^2(k+p)^2}.
\end{equation}

The diagram (6) makes the vanishing contribution, because it contains the block

\begin{equation}
\frac{(D_1^+)^4 (D_2^+)^4}{\Box} \delta^{14}(z_1-z_2) \frac{(u_1^- u_2^-)}{(u_1^+ u_2^+)^3}\Bigg|_{z_1=z_2;\,u_1=u_2} = (u_1^+ u_2^+) (u_1^- u_2^-) \frac{1}{\Box}\delta^6(x_1-x_2)\Big|_{u_1=u_2} = 0.
\end{equation}

The contribution of the Nielsen--Kallosh ghosts $\varphi$ is given by the diagrams (4) and (7) in Fig. \ref{Figure_Gauge_Harmonic_Diagrams}.
Expressions for them differ from those corresponding to diagrams (3) and (6) by the factor $-1/2$.

However, the Nielsen--Kallosh ghost contribution also includes $\mbox{Det}^{1/2} \stackrel{\bm{\frown}}{\bm{\Box}}$, which is calculated
using the definition (\ref{Determinant}). It is easy to show that this determinant makes the vanishing contribution. Actually, the tadpole diagram (7)
contains $(D^+)^4\delta(z_1-z_2)\Big|_{z_1=z_2} = 0$, while the diagram of the type (4) contains $(u_1^+ u_2^+)\Big|_{u_1=u_2} = 0\,$.
Therefore, no contribution comes from the superfields $\xi^{(+4)}$ and $\sigma$ at all.

Thus, the total ghost contribution to the considered part of the one-loop effective action can be written as

\begin{equation}
\frac{i}{2} C_2 \int \frac{d^6p}{(2\pi)^6}\,  d^8\theta\, du_1\, du_2\, \bm{V}^{++ A}(-p,\theta,u_1)
\bm{V}^{++ A}(p,\theta,u_2) \frac{1}{(u_1^+u_2^+)^2} \int \frac{d^6k}{(2\pi)^6}\, \frac{1}{k^2(k+p)^2}.
\end{equation}

\subsection*{A2. Diagrams with the loop of the gauge superfield}
\hspace{\parindent}

The contribution containing a loop of the quantum gauge superfield
is given by the sum of the diagrams (2) and (5) in Fig. \ref{Figure_Gauge_Harmonic_Diagrams}. Let us start with the diagram (2).
It is convenient to split this diagram into the three parts. The first one contains two vertices coming from the classical action $S_{\mbox{\scriptsize SYM}}$,
the second one contains one vertex from $S_{\mbox{\scriptsize SYM}}$ and one vertex from the gauge-fixing action $S_{\mbox{\scriptsize gf}}$,
and the last one contains two vertices, both coming from $S_{\mbox{\scriptsize gf}}$.

We start with a sub-diagram containing two $S_{\mbox{\scriptsize SYM}}$-vertices. The Feynman rules give for it the following analytical expression:

\begin{eqnarray}
&& -\frac{i}{4} \int d^{14}z_1\, d^{14}z_2\,du_1\, du_2\,du_3\, du_4\,du_5\,du_6\, f^{ABC} f^{ABD} \bm{V}^{++ C}(z_1,u_3) \bm{V}^{++ D}(z_2,u_6)\nonumber\\
&& \times \frac{1}{(u_1^+u_2^+) (u_2^+u_3^+) (u_3^+ u_1^+) (u_4^+u_5^+) (u_5^+u_6^+) (u_6^+ u_4^+)}\, \frac{(D_1^+)^4}{\Box}\delta^{14}(z_1-z_2)
\delta^{(-2,2)}(u_1,u_4)\nonumber\\
&&\times \frac{(D_2^+)^4}{\Box}\delta^{14}(z_1-z_2) \delta^{(-2,2)}(u_2,u_5).
\end{eqnarray}

\noindent
First, we use the identity (\ref{Theta_Identity}) and calculate one of the $\theta$-integrals.
Then we calculate two harmonic integrals by making use of the harmonic $\delta$-functions. After this we obtain

\begin{eqnarray}
&& -\frac{i}{4} C_2 \int d^{6}x_1\, d^{6}x_2\,d^8\theta \,du_1\, du_2\,du_3\, du_6\, \bm{V}^{++ A}(x_1,\theta,u_3) \bm{V}^{++ A}(x_2,\theta,u_6)\nonumber\\
&&\times \frac{(u_1^+u_2^+)^2}{ (u_2^+u_3^+) (u_3^+ u_1^+) (u_2^+u_6^+) (u_6^+ u_1^+)}\, \frac{1}{\Box}\delta^{6}(x_1-x_2)\, \frac{1}{\Box}\delta^{6}(x_1-x_2).
\end{eqnarray}

\noindent
Then, we use the identity

\begin{equation}
(u_1^+ u_2^+)^2 = D_1^{++}\Big[(u_1^- u_2^+) (u_1^+ u_2^+)\Big]
\end{equation}

\noindent
in the numerator of the harmonic factor and integrate by parts with respect to $D_1^{++}$,  taking into account the relation (\ref{U_Identity}).
The resulting expression, written in the momentum representation, has the form

\begin{eqnarray}
&& \frac{i}{4} C_2 \int \frac{d^6p}{(2\pi)^6}\,  d^8\theta \,du_1\, du_2\,du_3\, du_6\, \bm{V}^{++ A}(-p,\theta,u_3) \bm{V}^{++ A}(p,\theta,u_6)
\int \frac{d^6k}{(2\pi)^6} \frac{1}{k^2 (k+p)^2}\nonumber\\
&&\times  \frac{(u_1^- u_2^+) (u_1^+ u_2^+)}{(u_2^+ u_3^+) (u_2^+ u_6^+)} \Big[\delta^{(1,-1)}(u_1,u_3) \frac{1}{(u_1^+ u_6^+)}
+ \delta^{(1,-1)}(u_1,u_6) \frac{1}{(u_1^+ u_3^+)}\Big].
\end{eqnarray}

\noindent
The harmonic $\delta$-functions allow one to do one of the harmonic integrals,

\begin{eqnarray}
&& -\frac{i}{2} C_2 \int \frac{d^6p}{(2\pi)^6}\,  d^8\theta \,du_1\, du_2\,du_3\, \bm{V}^{++ A}(-p,\theta,u_3) \bm{V}^{++ A}(p,\theta,u_1) \int \frac{d^6k}{(2\pi)^6} \frac{1}{k^2 (k+p)^2}\nonumber\\
&&\times  \frac{D_2^{++}(u_1^- u_2^-)}{(u_2^+ u_3^+) (u_1^+ u_3^+)},
\end{eqnarray}

\noindent
where we used the identity $(u_1^- u_2^+)=D_2^{++}(u_1^- u_2^-)$. Integrating by parts with respect to $D_2^{++}$,
we obtain the final expression for the considered part of the diagram (2),

\begin{equation}\label{SD1}
\frac{i}{2} C_2 \int \frac{d^6p}{(2\pi)^6}\,  d^8\theta \,du_1\,du_3\, \bm{V}^{++ A}(-p,\theta,u_3)
\bm{V}^{++ A}(p,\theta,u_1) \frac{(u_1^- u_3^-)}{(u_1^+ u_3^+)} \int \frac{d^6k}{(2\pi)^6} \frac{1}{k^2 (k+p)^2}
\end{equation}

The second sub-diagram (in the diagram (2)) contains one vertex coming from $S_{\mbox{\scriptsize SYM}}$ and another one coming from $S_{\mbox{\scriptsize gf}}$.
The corresponding expression reads

\begin{eqnarray}
&& -\frac{i}{2} C_2 \int d^{14}z_1\, d^{14}z_2\,du_1\, du_2\,du_3\, du_4\,du_5\, \Big[\bm{b}^{A}(z_1,u_1) - \bm{b}^{A}(z_1,u_2)\Big] \bm{V}^{++ A}(z_2,u_3)\nonumber\\
&& \times D_1^{++}\Big[\frac{(u_1^- u_2^+)}{(u_1^+ u_2^+)^3}\Big] \frac{1}{(u_3^+u_4^+) (u_4^+u_5^+) (u_5^+ u_3^+)}\, \frac{(D_1^+)^4}{\Box}\delta^{14}(z_1-z_2)
\delta^{(-2,2)}(u_1,u_4)\nonumber\\
&&\times \frac{(D_2^+)^4}{\Box}\delta^{14}(z_1-z_2) \delta^{(-2,2)}(u_2,u_5).
\end{eqnarray}

\noindent
We start the calculation, using the identity (\ref{Theta_Identity}). Then we calculate the integral over $d^8\theta_2$ by exploiting $\delta^8(\theta_1-\theta_2)$
and two harmonic integrals with the help of the harmonic $\delta$-functions present in the propagators. The  expression considered takes the form

\begin{eqnarray}
&& -\frac{i}{2} C_2 \int d^{6}x_1\, d^{6}x_2\, d^8\theta\, du_1\, du_2\,du_3\, \Big[\bm{b}^{A}(x_1,\theta,u_1) - \bm{b}^{A}(x_1,\theta, u_2)\Big] \bm{V}^{++ A}(x_2,\theta,u_3)\nonumber\\
&& \times D_1^{++}\Big[\frac{(u_1^- u_2^+)}{(u_1^+ u_2^+)^3}\Big] \frac{(u_1^+ u_2^+)^3}{(u_3^+u_1^+) (u_2^+ u_3^+)} \, \frac{1}{\Box}\delta^{6}(x_1-x_2)
\, \frac{1}{\Box}\delta^{6}(x_1-x_2).
\end{eqnarray}

\noindent
Calculating the action of harmonic derivative $D_1^{++}$ and passing to the momentum representation, we obtain

\begin{eqnarray}
&& -\frac{i}{2} C_2 \int \frac{d^6p}{(2\pi)^6}\,  d^8\theta\,du_1\, du_2\,du_3\, \Big[\bm{b}^{A}(-p,\theta,u_1) - \bm{b}^{A}(-p,\theta,u_2)\Big]
\bm{V}^{++ A}(p,\theta,u_3)\nonumber\\
&& \times \frac{(u_1^+ u_2^+)}{(u_3^+ u_1^+) (u_2^+ u_3^+)} \int \frac{d^6k}{(2\pi)^6} \frac{1}{k^2 (k+p)^2}.
\end{eqnarray}

\noindent
It is convenient to rewrite the numerator of the harmonic factor in this expression as

\begin{equation}
(u_1^+ u_2^+) = D_1^{++}(u_1^- u_2^+)
\end{equation}

\noindent
and then to integrate by parts with respect to $D_1^{++}$. Next, resorting to Eq. (\ref{U_Identity}), we find

\begin{eqnarray}
&& \frac{i}{2} C_2 \int \frac{d^6p}{(2\pi)^6}\,  d^8\theta\,du_1\, du_2\,du_3\,\int \frac{d^6k}{(2\pi)^6} \frac{1}{k^2 (k+p)^2}
\bm{V}^{++ A}(p,\theta,u_3) \Bigg[D_1^{++}\bm{b}^{A}(-p,\theta,u_1) \nonumber\\
&& \times \frac{(u_1^- u_2^+)}{(u_3^+ u_1^+) (u_2^+ u_3^+)} - \Big[\bm{b}^{A}(-p,\theta,u_1)
- \bm{b}^{A}(-p,\theta,u_2)\Big] \frac{(u_1^- u_2^+)}{(u_2^+ u_3^+)}\delta^{(1,-1)}(u_1,u_3) \Bigg].
\end{eqnarray}

\noindent
In the first term the derivative of the bridge gives the superfield $\bm{V}^{++}$, while in the second one
it is possible to take off one of the harmonic integrals,

\begin{eqnarray}
&&\hspace*{-12mm} -\frac{i}{2} C_2 \int \frac{d^6p}{(2\pi)^6}\,  d^8\theta\,\int \frac{d^6k}{(2\pi)^6}
\frac{1}{k^2 (k+p)^2}  \Bigg[ \int du_1\, du_2\,du_3\, \bm{V}^{++A}(-p,\theta,u_1) \bm{V}^{++ A}(p,\theta,u_3) \nonumber\\
&&\hspace*{-12mm} \times \frac{(u_1^- u_2^+)}{(u_3^+ u_1^+) (u_2^+ u_3^+)} + \int du_1\,du_2\,
\Big[\bm{b}^{A}(-p,\theta,u_1) - \bm{b}^{A}(-p,\theta,u_2)\Big] \bm{V}^{++ A}(p,\theta,u_1) \frac{(u_1^- u_2^+)}{(u_2^+ u_1^+)} \Bigg].
\end{eqnarray}

\noindent
Using the identity

\begin{equation}
(u_1^- u_2^+) = D_2^{++}(u_1^- u_2^-)
\end{equation}

\noindent
and integrating by parts with respect to $D_2^{++}$, this expression can be rewritten as

\begin{eqnarray}
&&\hspace*{-8mm} \frac{i}{2} C_2 \int \frac{d^6p}{(2\pi)^6}\,  d^8\theta\,\int \frac{d^6k}{(2\pi)^6} \frac{1}{k^2 (k+p)^2}
\Bigg[ \int du_1\, du_2\,du_3\, \bm{V}^{++A}(-p,\theta,u_1) \bm{V}^{++ A}(p,\theta,u_3) \nonumber\\
&&\hspace*{-8mm} \times \frac{(u_1^- u_2^-)}{(u_3^+ u_1^+)}\delta^{(1,-1)}(u_2,u_3)
- \int du_1\,du_2\, D_2^{++}\bm{b}^{A}(-p,\theta,u_2) \bm{V}^{++ A}(p,\theta,u_1) \frac{(u_1^- u_2^-)}{(u_2^+ u_1^+)} \Bigg].
\end{eqnarray}

\noindent
This implies that the contribution of the sub-diagram considered is finally given by the expression

\begin{equation}\label{SD2}
- i C_2 \int \frac{d^6p}{(2\pi)^6}\,  d^8\theta\, du_1\, du_3\,\int \frac{d^6k}{(2\pi)^6} \frac{1}{k^2 (k+p)^2}   \bm{V}^{++A}(-p,\theta,u_1)
\bm{V}^{++ A}(p,\theta,u_3)\frac{(u_1^- u_3^-)}{(u_1^+ u_3^+)}.
\end{equation}

The last sub-diagram contains two vertices coming from $S_{\mbox{\scriptsize gf}}$. It is written in the form

\begin{eqnarray}
&& -\frac{i}{4} C_2 \int d^{14}z_1\, d^{14}z_2\,du_1\, du_2\,du_3\, du_4\,\Big[\bm{b}^{A}(z_1,u_1)
- \bm{b}^{A}(z_1,u_2)\Big] \Big[\bm{b}^{A}(z_2,u_3) - \bm{b}^{A}(z_2,u_4)\Big]\nonumber\\
&& \times D_1^{++}\Big[\frac{(u_1^- u_2^+)}{(u_1^+ u_2^+)^3}\Big]
D_3^{++}\Big[\frac{(u_3^- u_4^+)}{(u_3^+ u_4^+)^3}\Big]  \frac{(D_1^+)^4}{\Box}\delta^{14}(z_1-z_2)
\delta^{(-2,2)}(u_1,u_3)\, \frac{(D_2^+)^4}{\Box}\delta^{14}(z_1-z_2) \nonumber\\
&&\times \delta^{(-2,2)}(u_2,u_4).
\end{eqnarray}

\noindent
As before, using Eq. (\ref{Theta_Identity}) and $\delta$-functions, we calculate one of the $\theta$-integrals and two harmonic integrals.
The result in the momentum representation is written as

\begin{eqnarray}
&& -\frac{i}{4} C_2 \int \frac{d^6p}{(2\pi)^6}\,  d^8\theta\,\int \frac{d^6k}{(2\pi)^6} \frac{1}{k^2 (k+p)^2}\int du_1\, du_2\,
\Big[\bm{b}^{A}(-p,\theta,u_1) - \bm{b}^{A}(-p,\theta,u_2)\Big] \nonumber\\
&& \times \Big[\bm{b}^{A}(p,\theta,u_1) - \bm{b}^{A}(p,\theta,u_2)\Big]
(u_1^+ u_2^+)^4 D_1^{++}\Big[\frac{(u_1^- u_2^+)}{(u_1^+ u_2^+)^3}\Big] D_1^{++}\Big[\frac{(u_1^- u_2^+)}{(u_1^+ u_1^+)^3}\Big].
\end{eqnarray}

\noindent
Calculating the harmonic derivatives in this expression, we obtain

\begin{eqnarray}
&& -\frac{i}{4} C_2 \int \frac{d^6p}{(2\pi)^6}\,  d^8\theta\,\int \frac{d^6k}{(2\pi)^6}
\frac{1}{k^2 (k+p)^2}\int du_1\, du_2\,\Big[\bm{b}^{A}(-p,\theta,u_1) - \bm{b}^{A}(-p,\theta,u_2)\Big]\nonumber\\
&&\times \Big[\bm{b}^{A}(p,\theta,u_1) - \bm{b}^{A}(p,\theta,u_2)\Big].
\end{eqnarray}

\noindent
Taking into account the relation

\begin{equation}
D_1^{++} D_2^{++}\Big[\frac{(u_1^- u_2^-)}{(u_1^+ u_2^+)}\Big] = D_1^{++} \Big[\frac{(u_1^- u_2^+)}{(u_1^+ u_2^+)}\Big] = 1 - \delta^{(1,-1)}(u_1,u_2),
\end{equation}

\noindent
the last expression can be equivalently written in the form

\begin{eqnarray}
&& -\frac{i}{4} C_2 \int \frac{d^6p}{(2\pi)^6}\,  d^8\theta\,\int \frac{d^6k}{(2\pi)^6} \frac{1}{k^2 (k+p)^2}\int du_1\, du_2\,\Big[\bm{b}^{A}(-p,\theta,u_1)
- \bm{b}^{A}(-p,\theta,u_2)\Big] \nonumber\\
&& \times \Big[\bm{b}^{A}(p,\theta,u_1) - \bm{b}^{A}(p,\theta,u_2)\Big] D_1^{++} D_2^{++}\Big[\frac{(u_1^- u_2^-)}{(u_1^+ u_2^+)}\Big].
\end{eqnarray}

\noindent
After integrating by parts with respect to the harmonic derivatives,  we obtain

\begin{equation}
\frac{i}{2} C_2 \int \frac{d^6p}{(2\pi)^6}\,  d^8\theta\,\int \frac{d^6k}{(2\pi)^6} \frac{1}{k^2 (k+p)^2}\int du_1\, du_2\,D_1^{++}
\bm{b}^{A}(-p,\theta,u_1) D_2^{++}\bm{b}^{A}(p,\theta,u_2) \frac{(u_1^- u_2^-)}{(u_1^+ u_2^+)},
\end{equation}

\noindent
that can be expressed in terms of $\bm{V}^{++}$ as

\begin{equation}\label{SD3}
\frac{i}{2} C_2 \int \frac{d^6p}{(2\pi)^6}\,  d^8\theta\,\int \frac{d^6k}{(2\pi)^6} \frac{1}{k^2 (k+p)^2}\int du_1\, du_2\,\bm{V}^{++A}(-p,\theta,u_1)
\bm{V}^{++A}(p,\theta,u_2)  \frac{(u_1^- u_2^-)}{(u_1^+ u_2^+)}.
\end{equation}

Summing up the contributions of the three sub-diagrams, (\ref{SD1}), (\ref{SD2}), and (\ref{SD3}), we conclude that the diagram (2) vanishes,

\begin{eqnarray}
&& \frac{i}{2} C_2 \int \frac{d^6p}{(2\pi)^6}\,  d^8\theta\,\int \frac{d^6k}{(2\pi)^6} \frac{1}{k^2 (k+p)^2}\int du_1\, du_2\,
\bm{V}^{++A}(-p,\theta,u_1) \bm{V}^{++A}(p,\theta,u_2)\nonumber\\
&&\times \frac{(u_1^- u_2^-)}{(u_1^+ u_2^+)}\Big(1-2+1\Big) = 0\,.
\end{eqnarray}

Diagram (5) is a sum of two parts. The first one contains the vertex coming from the SYM action (\ref{Action_N2SYM}),
while in the second part the vertex originates from the gauge-fixing term (\ref{GF_Term}). It is easy to see that the contribution of every part
is  vanishing  separately. In particular, the first part is proportional to

\begin{eqnarray}
&& C_2 \int d^{14}z\, du_1\, du_2\, du_3\, du_4\, \bm{V}^{++A}(z,u_1) \bm{V}^{++A}(z,u_3) \frac{1}{(u_1^+ u_2^+)(u_2^+ u_3^+) (u_3^+ u_4^+) (u_4^+ u_1^+)}\qquad \nonumber\\
&&\times (D_2^+)^4 \delta^{14}(z_1-z_2) \delta^{(-2,2)}(u_2,u_4)\Bigg|_{z_1=z_2=z}.
\end{eqnarray}

\noindent
This expression vanishes, because only four spinor derivatives are left there to act on the Grassmann $\delta$-function at coincident points.
The second part vanishes for the same reason.

Thus, the sum of both diagrams with the loop of the quantum gauge superfield inside yields zero.

\renewcommand\theequation{B.\arabic{equation}} \setcounter{equation}0
\section*{Appendix B\quad}

\section*{Gauge-hypermultiplet three-point function}

\subsection*{B1. Abelian case}
\hspace{\parindent}\label{Appendix_3Point_Abelian}

In this section we describe details of calculating the gauge-hypermultiplet three-point function in the abelian case. For the abelian theory only the left diagram
in Fig. \ref{Figure_One-Loop_Vertex} remains, while the right one is absent. In the calculations we use the minimal gauge $\xi_0=1$. Then the considered
contribution to the effective action of the abelian theory (\ref{Action_Abelian}) has the form

\begin{eqnarray}\label{Three-Point}
&&\hspace*{-5mm} - 2f_0^2 \int d\zeta^{(-4)}_1\,du_1\,d\zeta^{(-4)}_2\,du_2\,d\zeta^{(-4)}_3\,du_3\,\widetilde q^+(z_1,u_1)
q^+(z_3,u_3)\bm{V}^{++}(z_2,u_2) \frac{(D_1^+)^4}{\Box} \delta^{(2,-2)}(u_3,u_1)\nonumber\\
&&\hspace*{-5mm}\times \delta^{14}(z_1-z_3)\, \frac{1}{(u_1^+ u_2^+)^3}\frac{(D_1^+)^4 (D_2^+)^4}{\Box}
\delta^{14}(z_1-z_2)\, \frac{1}{(u_2^+ u_3^+)^3}\frac{(D_2^+)^4 (D_3^+)^4}{\Box} \delta^{14}(z_2-z_3).
\end{eqnarray}

\noindent
Like in the previous cases, we start by converting the integrals over $d\zeta^{(-4)}$ into integrals over the full measure $d^{14}z$,

\begin{eqnarray}
&& - 2f_0^2 \int d^{14}z_1\,du_1\,d^{14}z_2\,du_2\,d^{14}z_3\,du_3\,\widetilde q^+(z_1,u_1) q^+(z_3,u_3)\bm{V}^{++}(z_2,u_2)
\frac{1}{\Box} \delta^{(2,-2)}(u_3,u_1)\nonumber\\
&& \times \delta^{14}(z_1-z_3)\, \frac{1}{(u_1^+ u_2^+)^3}\frac{(D_1^+)^4 (D_2^+)^4}{\Box} \delta^{14}(z_1-z_2)
\, \frac{1}{(u_2^+ u_3^+)^3}\frac{1}{\Box} \delta^{14}(z_2-z_3).
\end{eqnarray}

\noindent
Using the $\delta$-functions, we can calculate integrals over $d^{8}\theta_3$ and $du_3$,

\begin{eqnarray}
&& 2f_0^2 \int d^{14}z_1\,du_1\,d^{14}z_2\,du_2\,d^{6}x_3\,\widetilde q^+(z_1,u_1) q^+(x_3,\theta_1,u_1)\bm{V}^{++}(z_2,u_2) \frac{1}{\Box} \nonumber\\
&& \times \delta^{6}(x_1-x_3)\, \frac{1}{(u_1^+ u_2^+)^6}\frac{(D_1^+)^4 (D_2^+)^4}{\Box} \delta^{14}(z_1-z_2)\, \frac{1}{\Box}
\delta^{6}(x_2-x_3)\delta^8(\theta_1-\theta_2).
\end{eqnarray}

\noindent
As the next step, using the identity (\ref{Theta_Identity}), we obtain

\begin{eqnarray}
&& 2f_0^2 \int d^{14}z_1\,du_1\,d^{14}z_2\,du_2\,d^{6}x_3\,\widetilde q^+(z_1,u_1) q^+(x_3,\theta_1,u_1)\bm{V}^{++}(z_2,u_2) \frac{1}{\Box} \nonumber\\
&& \times \delta^{6}(x_1-x_3)\, \frac{1}{(u_1^+ u_2^+)^2}\frac{1}{\Box} \delta^{14}(z_1-z_2)\, \frac{1}{\Box} \delta^{6}(x_2-x_3).
\end{eqnarray}

\noindent
The remaining $\delta$-function $\delta^8(\theta_1-\theta_2)$ can be used to perform the integration over $d^8\theta_2$.
The result of these manipulations in the coordinate representation is

\begin{eqnarray}
&& 2f_0^2 \int d^{6}x_1\,d^{6}x_2\,d^{6}x_3\,d^8\theta\,du_1\,du_2\,\widetilde q^+(x_1,\theta,u_1) q^+(x_3,\theta,u_1)\bm{V}^{++}(x_2,\theta,u_2) \frac{1}{\Box} \nonumber\\
&& \times \delta^{6}(x_1-x_3)\, \frac{1}{(u_1^+ u_2^+)^2}\frac{1}{\Box} \delta^{6}(x_1-x_2)\, \frac{1}{\Box} \delta^{6}(x_2-x_3).
\end{eqnarray}

\noindent
After converting this expression to the momentum representation (in the Minkowski space) we arrive at the following final answer

\begin{eqnarray}
&& -2f_0^2 \int \frac{d^{6}p}{(2\pi)^6}\,\frac{d^{6}q}{(2\pi)^6}\,d^8\theta\,du_1\,du_2\,\widetilde q^+(q+p,\theta,u_1) q^+(-q,\theta,u_1)\bm{V}^{++}(-p,\theta,u_2)
\frac{1}{(u_1^+ u_2^+)^2}\nonumber\\
&& \times\int \frac{d^{6}k}{(2\pi)^6}\, \frac{1}{k^2 (q+k)^2 (q+k+p)^2}.
\end{eqnarray}
\vspace{0.2cm}

\subsection*{B2. Non-abelian theory: The second diagram in Fig. \ref{Figure_One-Loop_Vertex}}
\hspace{\parindent}\label{Appendix_3Point_Non-Abelian}

In this section we outline the calculation of the diagram (2) in Fig. \ref{Figure_One-Loop_Vertex}.
It is convenient to split it into two pieces. The first one corresponds to that part of the three-point gauge vertex which comes from the classical action $S$ ,
while the second piece corresponds to that part of the vertex which comes from the gauge-fixing term $S_{\mbox{\scriptsize gf}}$.
We calculate these two contributions separately. The expression for the first piece constructed by the Feynman rules in harmonic superspace
is written as

\begin{eqnarray}
&& - 2i f_0^2 \int d\zeta^{(-4)}_1\, d\zeta^{(-4)}_2\, d^{14}z_3\,  du_1\, du_2\, du_3\, du_4\, du_5\, \frac{1}{(u_3^+ u_4^+)
(u_4^+ u_5^+) (u_5^+ u_3^+)} f^{ABC}\bm{V}^{++ C}(z_3,u_5)\nonumber\\
&&\times \widetilde q^+(z_1,u_1)^i (T^A T^B)_i{}^j q^+(z_2,u_2)_j \, \frac{1}{(u_1^+ u_2^+)^3} \frac{(D_1^+)^4 (D_2^+)^4}{\Box} \delta^{14}(z_1-z_2)
\, \frac{(D_1^+)^4}{\Box} \delta^{14}(z_1-z_3)\nonumber\\
&&\times \delta^{(2,-2)}(u_3,u_1) \, \frac{(D^+_2)^4}{\Box} \delta^{14}(z_2-z_3) \delta^{(2,-2)}(u_4,u_2).
\end{eqnarray}

\noindent
Once again, we start by converting the integrals over $d\zeta^{(-4)}$ into those  over $d^{14}z$. Also we use the identity

\begin{equation}
2 f^{ABC} T^A T^B = f^{ABC} [T^A, T^B] = i C_2 T^C\,.
\end{equation}

\noindent
Then the considered part of the diagram (2) is written as

\begin{eqnarray}
&& f_0^2 C_2 \int d^{14}z_1\, d^{14}z_2\, d^{14}z_3\, du_1\, du_2\, du_3\, du_4\, du_5\, \frac{1}{(u_3^+ u_4^+)
(u_4^+ u_5^+) (u_5^+ u_3^+)} \bm{V}^{++ C}(z_3,u_5)\nonumber\\
&&\times \widetilde q^+(z_1,u_1)^i (T^C)_i{}^j q^+(z_2,u_2)_j \, \frac{1}{(u_1^+ u_2^+)^3} \frac{(D_1^+)^4 (D_2^+)^4}{\Box}
\delta^{14}(z_1-z_2) \, \frac{1}{\Box} \delta^{14}(z_1-z_3)\nonumber\\
&&\times \delta^{(2,-2)}(u_3,u_1) \, \frac{1}{\Box} \delta^{14}(z_2-z_3) \delta^{(2,-2)}(u_4,u_2).
\end{eqnarray}

\noindent
Harmonic $\delta$-functions can be used to do two harmonic integrations,

\begin{eqnarray}
&& f_0^2 C_2 \int d^{14}z_1\, d^{14}z_2\, d^{14}z_3\, du_1\, du_2\, du_5\, \frac{1}{(u_2^+ u_5^+) (u_5^+ u_1^+)}
\widetilde q^+(z_1,u_1)^i \bm{V}^{++}(z_3,u_5)_i{}^j q^+(z_2,u_2)_j\nonumber\\
&& \times \frac{1}{(u_1^+ u_2^+)^4} \frac{(D_1^+)^4 (D_2^+)^4}{\Box} \delta^{14}(z_1-z_2) \, \frac{1}{\Box} \delta^{14}(z_1-z_3) \, \frac{1}{\Box} \delta^{14}(z_2-z_3).
\end{eqnarray}

\noindent
Two $\theta$-integrals can be calculated by using the $\delta$-functions and the identity (\ref{Theta_Identity}). This gives

\begin{eqnarray}
&& f_0^2 C_2 \int d^{6}x_1\, d^{6}x_2\, d^{6}x_3\,d^8\theta\, du_1\, du_2\, du_5\,
\widetilde q^+(x_1,\theta,u_1)^i \bm{V}^{++}(x_3,\theta,u_5)_i{}^j q^+(x_2,\theta,u_2)_j\nonumber\\
&& \times \frac{1}{(u_2^+ u_5^+) (u_5^+ u_1^+)}\, \frac{1}{\Box} \delta^{14}(z_1-z_2) \, \frac{1}{\Box} \delta^{14}(z_1-z_3) \, \frac{1}{\Box} \delta^{14}(z_2-z_3).
\end{eqnarray}

\noindent
After relabeling the integration variable as $u_5\to u_3$, this expression can be written in the momentum representation (in the Minkowski space)  as

\begin{eqnarray}
&& - f_0^2 C_2 \int \frac{d^6p}{(2\pi)^6} \frac{d^6q}{(2\pi)^6} \frac{d^6k}{(2\pi)^6} d^8\theta\,du_1\,du_2\,du_3\, \widetilde q^+(q+p,\theta,u_1)^i
\bm{V}^{++}(-p,\theta,u_3)_i{}^j q^+(-q,\theta,u_2)_j\nonumber\\
&& \times \frac{1}{k^2 (k+p)^2 (k+q+p)^2}\, \frac{1}{(u_2^+ u_3^+) (u_3^+ u_1^+)}.
\end{eqnarray}

\noindent
Let us now express the gauge superfield $\bm{V}^{++}$ through the bridge $b$ in the linearized approximation,

\begin{equation}
\bm{V}^{++}=-D^{++} \bm{b} + \mbox{irrelevant terms},
\end{equation}

\noindent
where we omitted all terms with higher degrees of the bridge superfield. This is justified, since we deal only with terms linear in the gauge superfield.
Then, after integration by parts, we obtain

\begin{eqnarray}
&&  f_0^2 C_2 \int \frac{d^6p}{(2\pi)^6} \frac{d^6q}{(2\pi)^6} \frac{d^6k}{(2\pi)^6} d^8\theta\,du_1\,du_2\,du_3\, \widetilde q^+(q+p,\theta,u_1)^i
\bm{b}(-p,\theta,u_3)_i{}^j q^+(-q,\theta,u_2)_j\nonumber\\
&& \times \frac{1}{k^2 (k+p)^2 (k+q+p)^2}\, D_3^{++}\Big[\frac{1}{(u_3^+ u_2^+) (u_3^+ u_1^+)}\Big].\label{9zero}
\end{eqnarray}

\noindent
According to the identity (\ref{U_Identity}), the derivative $D_3^{++}$ produces two harmonic $\delta$-functions, so \p{9zero} becomes

\begin{eqnarray}
&&  f_0^2 C_2 \int \frac{d^6p}{(2\pi)^6} \frac{d^6q}{(2\pi)^6} \frac{d^6k}{(2\pi)^6} d^8\theta\,du_1\,du_2\,du_3\,
\widetilde q^+(q+p,\theta,u_1)^i \bm{b}(-p,\theta,u_3)_i{}^j q^+(-q,\theta,u_2)_j\nonumber\\
&& \times \frac{1}{k^2 (k+p)^2 (k+q+p)^2} \Big[\delta^{(1,-1)}(u_3,u_2)\frac{1}{(u_3^+ u_1^+)} + \delta^{(1,-1)}(u_3,u_1)\frac{1}{(u_3^+ u_2^+)}\Big].
\end{eqnarray}

\noindent
Thus, the considered part of the diagram (2) can be finally written as

\begin{eqnarray}
&&  f_0^2 C_2 \int \frac{d^6p}{(2\pi)^6} \frac{d^6q}{(2\pi)^6} \frac{d^6k}{(2\pi)^6} d^8\theta\,du_1\,du_2\,du_3\, \widetilde q^+(q+p,\theta,u_1)^i
[\bm{b}(-p,\theta,u_1) - \bm{b}(-p,\theta,u_2)]_i{}^j \nonumber\\
&& \times q^+(-q,\theta,u_2)_j\, \frac{1}{k^2 (k+p)^2 (k+q+p)^2}\, \frac{1}{(u_1^+ u_2^+)}.
\end{eqnarray}

Now, let us turn to calculating that part of the diagram (2) in which the purely gauge vertex comes from the gauge-fixing term $S_{\mbox{\scriptsize gf}}$. This contribution
is written as

\begin{eqnarray}
&& 2i f_0^2 \int d\zeta^{(-4)}_1 d\zeta^{(-4)}_2\, d^{14}z_3\,  du_1\, du_2\, du_3\, du_4\, \widetilde q^+(z_1,u_1)^i T^A T^B q^+(z_2,u_2)_j f^{ABC} \nonumber\\
&& \times \big[\bm{b}^C(z_3,u_4)-\bm{b}^C(z_3,u_3)\big] D_3^{++} D_4^{++}\Big[\frac{(u_3^- u_4^-)}{(u_3^+ u_4^+)^3}\Big]\, \frac{(D_1^+)^4 (D_2^+)^4}{\Box} \delta^{14}(z_1-z_2) \frac{1}{(u_1^+ u_2^+)^3}\qquad\nonumber\\
&&\times \frac{(D_1^+)^4}{\Box} \delta^{14}(z_1-z_3)\, \delta^{(2,-2)}(u_3,u_1)\, \frac{(D_2^+)^4}{\Box} \delta^{14}(z_2-z_3)\, \delta^{(2,-2)}(u_4,u_2).
\end{eqnarray}

\noindent
Following the same strategy as for the diagrams handled before, we convert integrations over $d\zeta^{(-4)}$ into integrations over $d^{14}z$
and calculate two harmonic integrals by making use of the harmonic $\delta$-functions. This gives

\begin{eqnarray}
&& - f_0^2 C_2 \int d^{14}z_1 d^{14}z_2\, d^{14}z_3\, du_1\, du_2\, \widetilde q^+(z_1,u_1)^i \big[\bm{b}(z_3,u_2)-\bm{b}(z_3,u_1)\big]_i{}^j q^+(z_2,u_2)_j \nonumber\\
&& \times  D_1^{++} D_2^{++}\Big[\frac{(u_1^- u_2^-)}{(u_1^+ u_2^+)^3}\Big]\, \frac{(D_1^+)^4 (D_2^+)^4}{\Box} \delta^{14}(z_1-z_2) \frac{1}{(u_1^+ u_2^+)^3}
\, \frac{1}{\Box} \delta^{14}(z_1-z_3) \, \frac{1}{\Box} \delta^{14}(z_2-z_3).\quad\nonumber\\
\end{eqnarray}

\noindent
Next, we need to calculate integrals over the anticommuting variables and to bring the result into the momentum representation. We obtain

\begin{eqnarray}
&&  f_0^2 C_2 \int \frac{d^6p}{(2\pi)^6} \frac{d^6q}{(2\pi)^6}\frac{d^6k}{(2\pi)^6} d^8\theta\, du_1\, du_2\, \widetilde q^+(q+p,\theta,u_1)^i
\big[\bm{b}(-p,\theta,u_2)-\bm{b}(-p,\theta,u_1)\big]_i{}^j  \qquad \nonumber\\
&& \times q^+(-q,\theta,u_2)_j \frac{1}{k^2 (k+p)^2 (k+p+q)^2} (u_1^+ u_2^+) D_1^{++} D_2^{++}\Big[\frac{(u_1^- u_2^-)}{(u_1^+ u_2^+)^3}\Big].
\end{eqnarray}

\noindent
Using the identity (\ref{U_Identity}) and the equality $(u_1^-u_2^+) \delta^{(2,-2)}(u_1,u_2)=-\delta^{(2,-2)}(u_1,u_2)\,$, this expression can be equivalently
represented as

\begin{eqnarray}
&&\hspace*{-5mm}  f_0^2 C_2 \int \frac{d^6p}{(2\pi)^6} \frac{d^6q}{(2\pi)^6}\frac{d^6k}{(2\pi)^6} d^8\theta\, du_1\, du_2\,
\widetilde q^+(q+p,\theta,u_1)^i \big[\bm{b}(-p,\theta,u_2)-\bm{b}(-p,\theta,u_1)\big]_i{}^j \qquad \nonumber\\
&&\hspace*{-5mm} \times q^+(-q,\theta,u_2)_j  \frac{1}{k^2 (k+p)^2 (k+p+q)^2} (u_1^+ u_2^+) \Big[\frac{1}{(u_1^+ u_2^+)^2}
- \frac{1}{2} (D_1^{--})^2 \delta^{(2,-2)}(u_1,u_2)\Big].
\end{eqnarray}

\noindent
Integrating  by parts in the last term with respect to $(D_1^{--})^2$, after some algebra we obtain

\begin{eqnarray}
&&  f_0^2 C_2 \int \frac{d^6p}{(2\pi)^6} \frac{d^6q}{(2\pi)^6}\frac{d^6k}{(2\pi)^6} d^8\theta\,  \frac{1}{k^2 (k+p)^2 (k+p+q)^2}  \Bigg[\int du_1\, du_2\,
\widetilde q^+(q+p,\theta,u_1)^i \nonumber\\
&& \times \big[\bm{b}(-p,\theta,u_2) -\bm{b}(-p,\theta,u_1)\big]_i{}^j q^+(-q,\theta,u_2)_j \frac{1}{(u_1^+ u_2^+)} - \int du\, \widetilde q^+(q+p,\theta,u)^i
\nonumber\\
&& \times D^{--}\bm{b}(-p,\theta,u)_i{}^j q^+(-q,\theta,u)_j \Bigg].
\end{eqnarray}

\noindent
The first term in this expression cancels the previous part of the considered diagram. Therefore, taking into account that

\begin{equation}
\bm{V}_{\mbox{\scriptsize linear}}^{--}=-D^{--} \bm{b} + \mbox{irrelevant terms},
\end{equation}

\noindent
the net result for the contribution of the diagram (2) in Fig. \ref{Figure_One-Loop_Vertex} is given by the expression

\begin{eqnarray}
&& f_0^2 C_2 \int \frac{d^6p}{(2\pi)^6} \frac{d^6q}{(2\pi)^6} \frac{d^6k}{(2\pi)^6} d^8\theta\,
\frac{1}{k^2 (k+p)^2 (k+p+q)^2} \int du\, \widetilde q^+(q+p,\theta,u)^i \bm{V}_{\mbox{\scriptsize linear}}^{--}(-p,\theta,u)_i{}^j \nonumber\\
&& \times q^+(-q,\theta,u)_j.
\end{eqnarray}

\end{document}